\documentclass[12pt]{article}
\pdfoutput=1

\newif\ifshowlabel   
\showlabelfalse  

\usepackage{amstext,amsmath,amsthm,amssymb,amsfonts,bbm,enumerate}
\usepackage[pagebackref, colorlinks = true, linkcolor = blue, urlcolor  = blue, citecolor = red]{hyperref}
\usepackage{url}
\usepackage[latin1]{inputenc}
\usepackage{hyperref}
\usepackage{amsthm}
\usepackage{color}
\usepackage{geometry}
\usepackage{graphicx}
\usepackage{framed}
\usepackage{mathbbol}
\usepackage{enumitem}

\allowdisplaybreaks
 \newcommand{\be}{\begin{equation}}
\newcommand{\ee}{\end{equation}}

\ifshowlabel
\usepackage[a3paper,landscape,margin=5cm]{geometry}
\usepackage{showkeys}
\fi

\def\CR{\nonumber \\}
\def\eq#1{(\ref{#1})}
\def\s[#1\s]{\begin{align}\begin{split}#1\end{split}\end{align}}
\def\[#1\]{\begin{align}#1\end{align}}

\newcommand{\cG}{{\mathcal G}}
\newcommand{\cZ}{{\mathcal Z}}
\newcommand{\cF}{{\mathcal F}}

\newcommand{\bG}{{\mathbb{G}}}
\newcommand{\bR}{{\mathbb{R}}}

\newcommand{\Tr}{\mathrm{Tr}}

\newcommand{\Fd}[1][]{{F_d^{#1}}}
\newcommand{\Fs}[1][]{{F_s^{#1}}}

\newcommand{\dom}{{\text{dom}}}

\begin{document}

\begin{titlepage}

\title{
\hfill\parbox{4cm}{ \normalsize YITP-19-17
}\\ 
\vspace{1cm} 
A random matrix model \\ with non-pairwise contracted indices}

\author{
Luca Lionni\footnote{luca.lionni@yukawa.kyoto-u.ac.jp },
Naoki Sasakura\footnote{sasakura@yukawa.kyoto-u.ac.jp}
\\
{\small{\it Yukawa Institute for Theoretical Physics, Kyoto University,}}
\\ {\small{\it  Kitashirakawa, Sakyo-ku, Kyoto 606-8502, Japan}}
}

\date{\today}

\maketitle

\begin{abstract}
We consider a random matrix model with both pairwise and non-pairwise contracted indices. The partition function of the matrix model is similar to that appearing in some replicated systems with random tensor couplings, such as the $p$-spin spherical model for the spin glass. 
We analyze the model using Feynman diagrammatic expansions, and provide an exhaustive characterization of the graphs which dominate when the dimensions of the pairwise and (or) non-pairwise contracted indices are large.
We apply this to investigate the properties of the wave function of a toy model closely related to a tensor model in the Hamilton formalism, which is studied in a quantum gravity context, and obtain a result in favor of the consistency of the quantum probabilistic interpretation of this tensor model.
\end{abstract}

\end{titlepage}

\tableofcontents

\section{Introduction}
\label{sec:model}
Random matrix models \cite{Wigner, thooft-planar, Brezin:1990rb,Douglas:1989ve,Gross:1989vs}  were first introduced by Wigner in a context of nuclear physics, and have since then proven to be an essential tool in modern physics and mathematics, with applications in quantum chromodynamics, disordered systems, 2D quantum gravity, quantum information,  combinatorics of discrete surfaces,  free probability, and so on \cite{2DgravityReview, oxford-rand-mat, book-rand-mat-2}. 

Our main interest in this paper is to study a new kind of random one-matrix model defined by the following partition function, 
\[
Z_{N,R}(\lambda, k) := \int_{\mathbb{R}^{NR}} d\phi \exp \left( -\lambda U(\phi)-k \Tr\phi\phi^t  \right),
\label{eq:integral}
\] 
where  the integration is done over matrices $\phi$ with real coefficients $\phi_a^i\in \mathbb{R}\ (a=1,2,\ldots,N,\ i=1,2,\ldots,R)$,  
$d \phi:=\prod_{i=1}^R \prod_{a=1}^N d\phi_a^i$,  the Gaussian part is $\Tr\phi\phi^t = \sum_{i=1}^R \sum_{a=1}^{N}\phi^i_a\phi_a^i$,  and the interaction term is 
\be
U(\phi) = \sum_{i,j=1}^R  \left(\sum_{a=1}^{N}\phi^i_a\phi^j_a \right)^3=  \sum_{a,b,c=1}^N \sum_{i,j=1}^R \phi_a^i \phi_b^i \phi_c^i \phi_a^j \phi_b^j \phi_c^j.
\label{eq:interaction}
\ee
The parameters $k,\lambda$ can be both real or complex, depending on the specific problems considered. 
 Random matrix models are usually defined using trace invariants and matrix products, for which the indices of the matrices are contracted (summed) pairwise. The archetypal example of one-matrix model is obtained for interactions of the form $\tilde U(\phi) = \Tr\bigl((\phi\phi^t)^p\bigr)$. 
Instead, while  the lower indices are contracted pairwise in the interaction we consider \eqref{eq:interaction},  the upper indices, $i,j\in\{1,2,\ldots,R\}$, do not appear pairwise. In our study, we will consider both square matrices ($R=N$) and rectangular matrices.  

Rectangular random matrix models were considered  and then systematically analyzed in 
\cite{Anderson:1990nw,Anderson:1991ku,Myers:1992dq}, extending the celebrated double scaling limits of 
matrix models \cite{Brezin:1990rb,Douglas:1989ve,Gross:1989vs}. See also 
\cite{Francesco-rect}  and references therein. 
In the large matrix size limit,  rectangular random matrix models interpolate between the behavior of branched polymers (involving Feynman graphs with a tree-like filamentary structure) and that of two-dimensional quantum gravity (involving planar Feynman ribbon graphs). 
An important step in solving these models was  to diagonalize the rectangular matrix by using the Lie-group symmetries 
on the matrix indices. On the other hand, the present model \eq{eq:integral} respects only the discrete permutation  symmetry\footnote{Namely, reordering of $i=\{1,2,\ldots,R\}$.} on the upper index, while it respects the orthogonal symmetry on the lower indices. 
Moreover, the usual pairwise contraction pattern allows for the t'Hooft expansion \cite{thooft-planar} over ribbon graphs, discrete surfaces classified according to their genera, where the contribution in the matrix sizes of a graph is given in terms of closed loops called faces \cite{2DgravityReview, Francesco-rect}. With the non-pairwise contraction pattern in \eqref{eq:interaction} we lose this combinatorial structure and the expansion over random discretized surfaces.
Because of the differences in the symmetry and combinatorial structure, 
we expect the present model \eq{eq:integral} to  behave differently from the usual square and rectangular random matrix models.
It is also challenging to analyze the present model with this lack of symmetry and without the topological expansion over discrete surfaces.

Due to this lack of symmetry, the present model, \eq{eq:integral} with \eq{eq:interaction},  can be seen as a random vector model with multiple vectors, $\phi^i\in \mathbb{R}^N\ (i=1,2,\ldots,R)$.  In the usual solvable settings of the vector models \cite{Nishigaki:1990sk,DiVecchia:1991vu}, however,  there are independent Lie-group symmetries for each vector, 
and the interactions are rather arbitrary among
the invariants made of these vectors.
On the other hand, our present model has more restrictive characteristics: there is only a single common Lie-group symmetry\footnote{Such a model was written down as (2.14) in the paper \cite{Nishigaki:1990sk}. However, the model was not solved.}, the vectors are equivalent with each other under the permutation symmetry, and the interaction has the particular form with  non-pairwise index contractions.
Therefore, we would expect that our model  defines a specific type of vector model with some interesting characteristic properties.
In a sense, the present model is in-between the matrix and vector models, and in fact, by just changing the power of the interaction term in \eq{eq:interaction} from 3 to 2, we recover the  usual $\Tr\bigl((\phi\phi^t)^2\bigr)$ rectangular random matrix model.

As a matter of fact, an expression very similar to \eq{eq:integral} has already been 
discussed in the context of spin glasses in physics, for the $p$-spin spherical model \cite{pspin, pedestrians}.
The model has spherical coordinates as degrees of freedom, and considers random couplings among them to model the spin glass.  An expression of the form \eq{eq:integral} appears after integrating out the random couplings under the replica trick.
However, there are some differences with our case:
there exists a constraint $\sum_{a=1}^N \phi_a^i  \phi_a^i=  const$, corresponding to spherical coordinates;
$\lambda$ is negative, while it should be positive  for the convergence of \eq{eq:integral} (or should have a positive real part); 
the limit $R\rightarrow 0$ is taken in applying the replica trick.
Because of these rather non-trivial differences, we would expect new outcomes with respect to the previous studies. Note that the case where $R$ is kept finite while $N$ is taken to be large in \eq{eq:integral} could have an application for systems with a finite number of ``real" replicas \cite{MezPar, pedestrians}. 

One of our motivations to initiate the study of the model \eq{eq:integral} is to investigate the properties of 
the wave function \cite{Narain:2014cya,Obster:2017dhx} of a tensor model \cite{Ambjorn:1990ge,Sasakura:1990fs,Godfrey:1990dt} 
in the Hamilton formalism \cite{Sasakura:2011sq,Sasakura:2012fb},
which is studied in a quantum gravity context. The expression \eq{eq:integral} can be obtained after integrating over the tensor argument of the wave function of the toy model introduced in  \cite{Obster:2017pdq}, which is closely related to this tensor model. The details will be explained in Section~\ref{sec:limit}. As another potential application, we can consider
randomly connected tensor networks \cite{Sasakura:2014zwa,Sasakura:2014yoa} with random tensors. It would also be possible to obtain \eq{eq:integral} by considering a random coupling vector model, or a bosonic timeless analogue of the SYK model \cite{SYKSY,SYKK}. Indeed, introducing $R$ replicas in such a model, we obtain 
\[
Z_{N,R}(\lambda, k) = \int dP e^{-\frac 1 2 \sum_{abc=1}^N P_{abc}^2}\Bigl(\int_{\bR^N}d\phi e^{ - k \sum_{a=1}^N \phi_a^2  - I\sqrt{2\lambda } \sum_{a,b,c=1}^N P_{abc} \phi_a \phi_b \phi_c }\Bigr)^R,
\] 
(where here the $\phi$ are vectors and $d\phi = \prod_{a=1}^N d\phi_a$) from which we recover \eq{eq:integral} by integrating out the random tensors. 
In fact, as detailed in this paper, the Feynman diagrammatic expansions of vector models with random couplings such as the SYK model with a finite number of replicas are still dominated by the celebrated melonic diagrams \cite{melons1, melons2, melons3} when the size of the system is large. We will show that this dominance still holds when the number of replicas is large, as long as the latter does not exceed the size of the system.\footnote{The results we obtain concerning dominant Feynman graphs should still apply to models with a time dependence.}

This paper is organized as follows. In Section~\ref{sec:graph}, we describe the Feynman graph expansion of  the partition function \eqref{eq:integral}. We identify the graphs for which the dependence in $N$ and $R$ is the strongest when the number of interactions is fixed in the following different regimes: $N$ large and $R$ finite, $R$ large and $N$ finite, and $R\sim N^\alpha$ with $\alpha \in (0, +\infty)$. 
In Section~\ref{sec:convergent},
we develop a method to treat
the model in a convergent series 
by separating the integration variables of \eq{eq:integral} into the angular and radial parts.
We apply the method to study 
the properties of the wave function of the toy
model introduced in \cite{Obster:2017pdq},
which is closely related to 
the tensor model mentioned above.
The last section is devoted to a summary and future prospects.

\section{Graphical expansion and dominant graphs  for the different regimes }
\label{sec:graph}

We consider the normalized partition function 
\be 
\label{eq:Part-Funct}
\cZ_{N,R}(\lambda, k) =\Bigl(\frac {k}{\pi}\Bigr)^{\frac{NR}2} \int_{\bR^{NR}} d\phi e^{-\lambda U(\phi) - k \Tr (\phi \phi^t)},
\ee 
where $\Tr(\phi \phi^t) = \sum_{a=1}^N \sum_{i=1}^R \phi_a^i\phi_a^i$, and where the interaction $U(\phi)$ is not an usual trace invariant, but instead has non-pairwise contracted indices,
\be 
\label{eq:Interaction}
U(\phi) = \sum_{i,j=1}^R \Bigl(\sum_{a=1}^N \phi_a^i \phi_a^j\Bigr)^3 =  \sum_{a,b,c=1}^N \sum_{i,j=1}^R \phi_a^i \phi_b^i \phi_c^i \phi_a^j \phi_b^j \phi_c^j.
\ee
This partition function is indeed normalized, as $\int_{\bR^{NR}} d\phi e^{ - k \Tr (\phi \phi^t)} = (\frac {\pi}{k})^{\frac{NR}2}$.

\begin{figure}[!h]
 \begin{center}
 \includegraphics[scale=0.7]{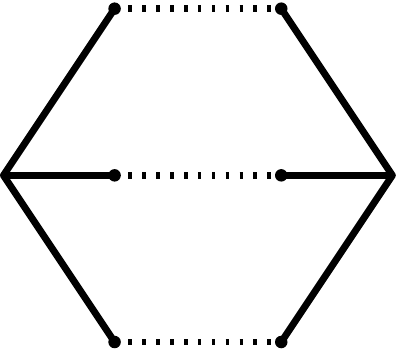}
 \caption{Graphical representation of an interaction  $\sum_{i,j=1}^R (\sum_{a=1}^N \phi^i_a\phi^j_a)^3$. 
 The variables $\phi_a^i$ are located at each connection point between 
 the solid and the dotted lines.
 The solid lines represent the contractions 
 of the $i$ and $j$ indices, while the dotted lines represent
 the contractions 
 of the lower indices.
  }
 \label{fig:intvertex}
 \end{center}
 \end{figure}

We represent graphically the contraction pattern of the interaction \eqref{eq:Interaction} in Fig.~\ref{fig:intvertex}. Each matrix $\phi$ is associated with a vertex, with two half-edges\footnote{An edge between two vertices is divided in two parts, which correspond to the neighborhoods of the two vertices. We call these parts half-edges.} attached: a dotted half-edge representing the lower index (summed from 1 to $N$), and a solid half-edge representing the upper index (summed from 1 to $R$). The  dotted half-edges are associated pairwise, representing the summation of the indices $a,b,c$ in \eqref{eq:Interaction}, while the solid half-edges are attached to trivalent nodes, representing the summation of the indices $i,j$ in  \eqref{eq:Interaction}. 

We consider the {\it formal} expansion of the partition function in powers of the coupling constant $\lambda$. It is formally obtained by expanding the exponential of the interaction \eqref{eq:Interaction} in \eqref{eq:Part-Funct}, by exchanging the sum and the integral, and by applying Wick theorem to compute Gaussian expectation values of products of \eqref{eq:Interaction} of the form
\be 
 \langle U(\phi) ^n \rangle_0  = \Bigl(\frac {k}{\pi}\Bigr)^{\frac{NR}2}  \int d\phi\, U(\phi) ^n e^{-k \Tr (\phi \phi^t)}. 
\ee  
This way, the partition function is formally expressed as 
\be 
\label{eq:Part-Func-Exp-1}
\cZ_{N,R}(\lambda, k) = \sum_{n \ge 0} z_n(N,R,k)(- \lambda)^n, \qquad  z_n(N,R,k) = \frac 1 {n!}  \langle U(\phi) ^n \rangle_0. 
\ee
By changing variables $\phi' = \sqrt{2k} \phi$,
we see that $z_n(N,R,k) = z'_n(N,R)/{(8k^3)^n}$.
In the present section, we identify the dominant term in $z_n(N,R,k)$  for different regimes of large $N$ and $R$.

\subsection{Feynman graphs}
\label{sub:Feyn}

\begin{figure}[!h]
 \begin{center}
 \includegraphics[scale=0.5]{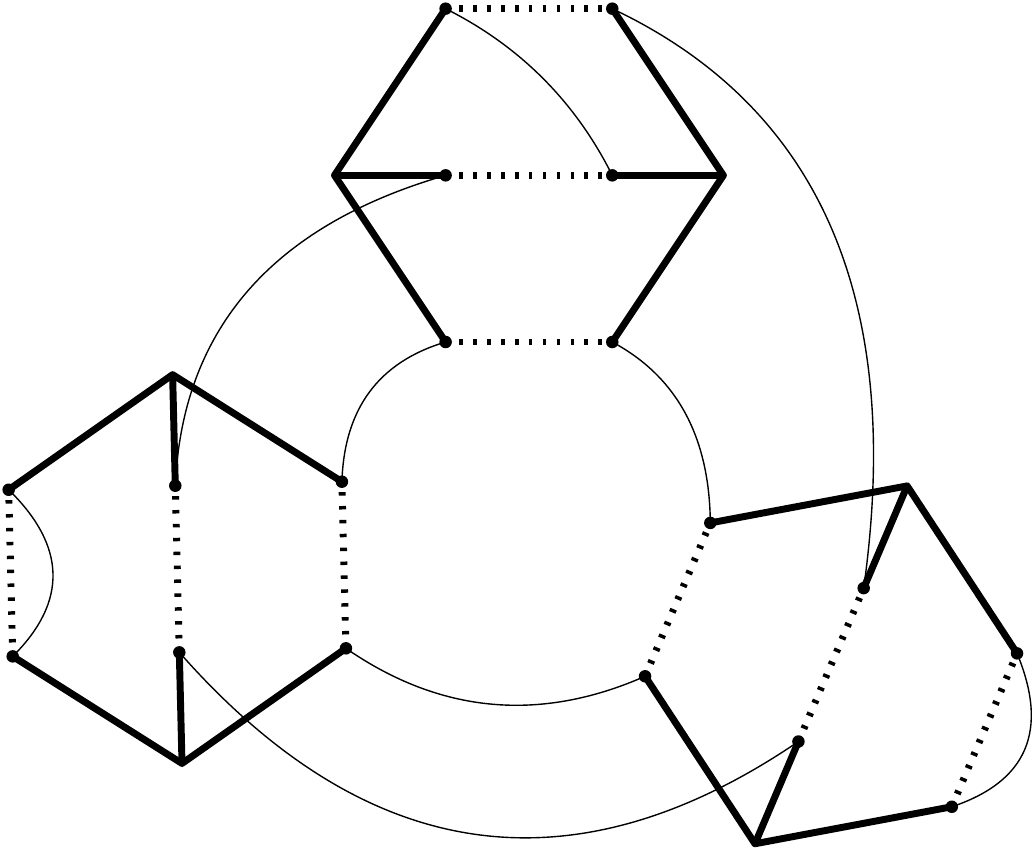}
 \caption{An example of a connected Feynman graph with  three interactions. Wick contractions are represented by the thin lines.}
 \label{fig:contraction}
 \end{center}
 \end{figure}

Applying Wick theorem, $\langle U(\phi) ^n \rangle_0$ is standardly 
computed by summing over all possible ways to pair the $6n$ matrices involved, and by replacing the paired matrices  with the Gaussian covariance 
\be 
\label{eq:eachwick}
\langle \phi^i_a \phi^j_b\rangle_0 = \frac 1 {2k} \delta_{ij}\delta_{ab}.
\ee 
This can be expressed graphically using sums over graphs as follows: the $n$ interactions $U(\phi)$ are each represented as in Fig.~\ref{fig:intvertex}, and contribute with a factor $(-\lambda)$, while the Wick pairings (the propagators) are represented by new thin edges between pairs of matrices, which identify the indices corresponding to the dotted and the solid edges, and contribute with a factor $1/2k$. We therefore have graphs with three kind of edges, dotted, solid and thin, and so that we recover  $n$  copies of the graph in Fig.~\ref{fig:intvertex} when the thin edges are deleted. We denote $\bG(n)$ the set of such graphs, and $\bG$ the set of graphs with any positive number of interactions. Similarly, we denote by $\bG_c(n)$ and $\bG_c$ the subsets of connected graphs in $\bG(n)$ and $\bG$. An example of a graph in $\bG_c(3)$ is represented in Fig.~\ref{fig:contraction}. 

\begin{figure}[!h]
 \begin{center}
 \includegraphics[scale=0.5]{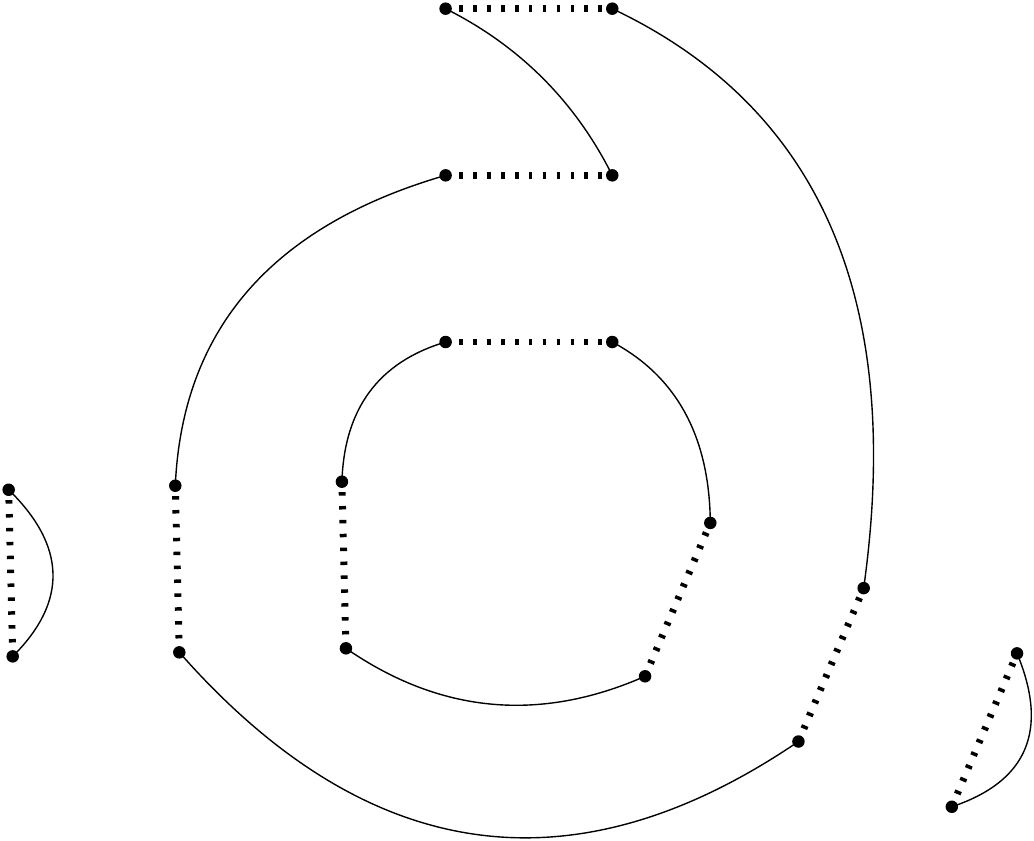}\hspace{2cm}\includegraphics[scale=0.5]{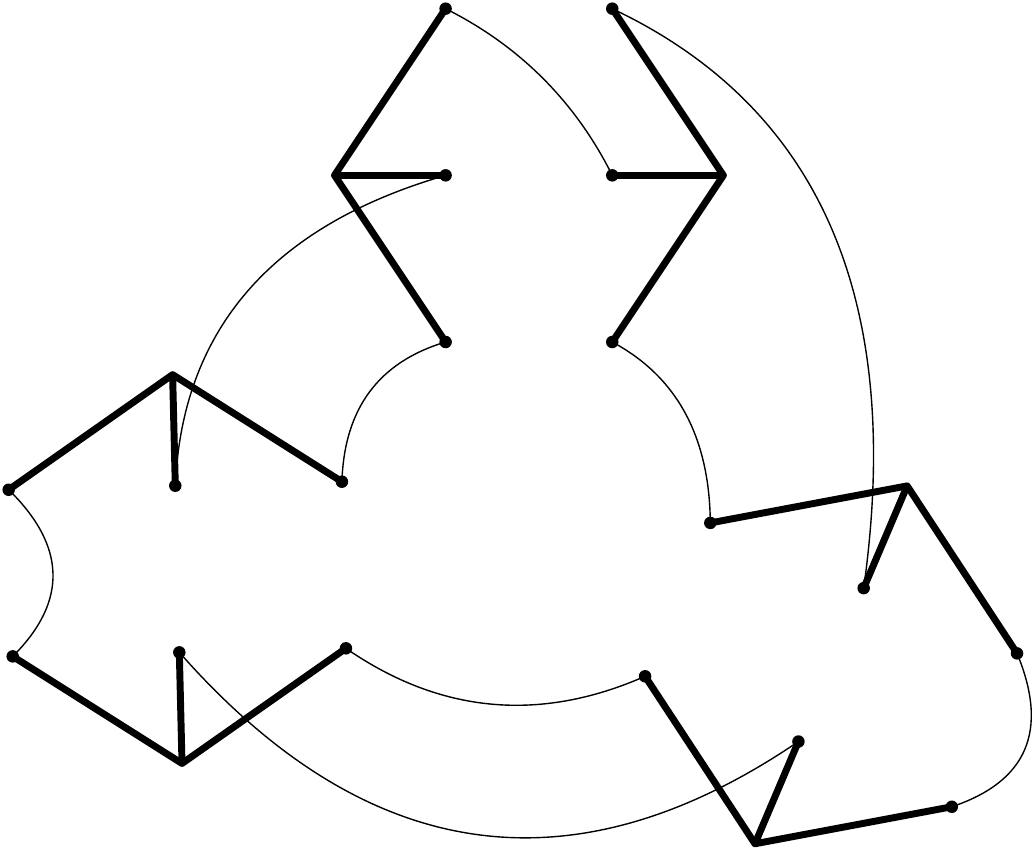} \caption{Faces for the graph of Fig.~\ref{fig:contraction}. On the left there are four dotted faces, representing the four free sums of the lower indices generated by the Wick contractions, each of which contributes with a factor of  $N$. There is a single solid face, represented on the right, thus one free sum for the upper indices, which generates a factor $R$. Hence, the total weight in $N,R$ of this Feynman graph is $N^4R$.}
 \label{fig:faces}
 \end{center}
 \end{figure}

  As for usual matrix models, the sums of Kronecker deltas corresponding to the indices contracted pairwise in \eqref{eq:Interaction} yield a factor of $N$ for each free sum on the lower index of the matrices. In our representation, these free sums correspond to
  the connected  subgraphs obtained when only the dotted and thin edges are kept, while the solid edges are deleted. These subgraphs are loops called \textit{dotted faces}\footnote{It is a common denomination in random matrix and tensor models to call such loops faces.}.  In Fig.~\ref{fig:contraction} for instance, there are four dotted faces, represented on the left of Fig.~\ref{fig:faces}, thus a contribution of $N^4$ for this graph. 
  
In the present case however, the contraction patterns of the upper indices corresponding to the solid edges are more complicated: we still get a factor of $R$ for every connected subgraph with only solid and thin edges, but now such subgraphs are no longer loops as they have nodes of valency three, as shown on the right of Fig.~\ref{fig:faces}.    Even though these subgraphs are not loops, we call them {\it solid faces}. 
  
  We denote by $\Fd$ (resp.~$\Fs$) the number of dotted (resp.~solid) faces. For the graph of Fig.~\ref{fig:contraction}, we thus have $\Fd=4$ and $\Fs=1$. Then, using Wick theorem, the expectation values $ \langle U(\phi) ^n \rangle_0$ are expressed as 
  \be 
 \langle U(\phi) ^n \rangle_0 =\Bigl(\frac{1 }{8k^3}\Bigr)^{n}   \sum_{G\in \bG(n)} m(G) N^{\Fd(G)} R^{\Fs(G)},
  \ee 
where we have used the fact that the number of thin edges is $3n$, and where $m(G)$ is the multiplicity of the graph $G$, defined as the number of occurrences of $G$ when adding the thin edges in all possible ways for the $n(G)$ interactions. Inserting this in \eqref{eq:Part-Func-Exp-1}, the partition function  is formally expressed as an expansion indexed by Feynman graphs. 

\

\begin{figure}[h!]
 \begin{center}
 \includegraphics[width=12cm]{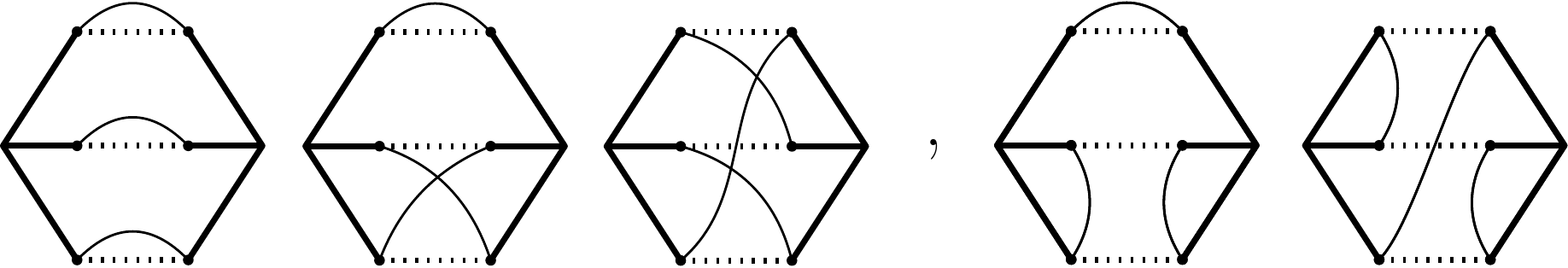}
 \caption{The list of Feynman graphs in $\bG(1)$. For later convenience,
 they are gathered in two groups, separated by a comma.}
 \label{fig:neq1}
 \end{center}
 \end{figure}
In Figure~\ref{fig:neq1}, all the possible one-interaction graphs  are drawn. From the graphs, one can 
easily compute the weights coming from the free sums over the indices.
One also has to take into account the multiplicities $m(G)$ of the graphs. For instance, 
the contribution of the first graph of Figure~\ref{fig:neq1} 
can be computed as $N^3 R(-\lambda/(2k)^3)$. By computing similarly for the
other graphs, the contributions of these graphs lead to
\[
z_1(N,R,k)=\left(N^3R+3 N^2 R+2 NR+3N^2R+6NR\right)(2k)^{-3}.
\] 
The terms are ordered in the same way as the graphs appear in Figure~\ref{fig:neq1}.

\

In practice, $\cZ_{N,R}$ is rather computed by exponentiating the \textit{free-energy}\footnote{In this paper, we call $\log {\cal Z}_{N,R}$ the free energy, rather than the real free energy $-\log {\cal Z}_{N,R}$ in physics to avoid the frequent appearance of extra minus signs. This ``convention" is often used in combinatorics papers.}
whose expansion involves only \textit{connected} Feynman graphs:   
  \be 
 \cF_{N,R}(\lambda, k) =  \log \cZ_{N,R}(\lambda, k) =\sum_{G\in \bG_c}  \frac {m(G)} {n(G)!} \Bigl(\frac{-\lambda}{8k^3}\Bigr)^{n(G)} N^{\Fd(G)} R^{\Fs(G)},
 \label{eq:free-nrj}
  \ee 
 where we have denoted by $n(G)$ the number of interactions $U(\phi)$ in the graph $G\in\bG_c$. 
 
 \

  \textit{
  Our aim in the present section, is to identify the graphs in $\bG_c(n)$ which dominate when $N$ or $R$ is large, or both, in various specific regimes.}

  \ 
  
  More precisely, we will consider the following cases: $N$ large and finite $R$ in Sec.~\ref{sub:LargeNfiniteR}, $R$ large and finite $N$ in Sec.~\ref{sub:LargeRfiniteN}, and both $N$ and $R$ large with $R\sim N^\alpha$, where $\alpha>1$ in Sec.~\ref{sub:AlphaLarg1} and where  $\alpha\le1$ in Sec.~\ref{sub:AlphaSmall1}. For each one of these regimes, and for a fixed value of $n\ge 1$, the connected graphs in $\bG_c(n)$ can be classified according to their dependence in $N$ and $R$. The graphs in $\bG_c(n)$ for which this dependence is the strongest are called \emph{dominant graphs}. We will compute the dominant free-energy, i.e.~the free energy restricted to dominant graphs. 
  
  \
  
  Note that because all the one-interaction graphs have the same contribution in $R$, in all the regimes where $N$ is large, the only dominant one-interaction graph is given by  the leftmost graph in Fig.~\ref{fig:neq1}, so that in any regime we may consider where $N$ is large, 
  \be 
  \label{eq:leading-order-1}
  z^\dom_1(N>>1,R,k) = \frac {N^3R}{8k^3}.
  \ee 
  
\begin{figure}[h!]
 \begin{center}
 \includegraphics[scale=0.8]{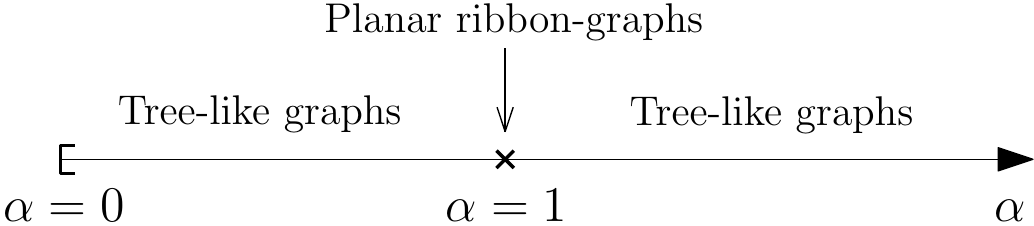}
 \caption{Dominant graphs at large $N$ for the random $N\times N^\alpha$ matrix models with $\alpha> 0$.}
 \label{fig:RecDomMat}
 \end{center}
 \end{figure}

  For the rectangular matrix model defined with an interaction of the form $\tilde U(\phi) = \Tr\bigl((\phi\phi^t)^p\bigr) $ with $p\ge 2$, the Feynman graphs are ribbon graphs whose vertices have $2p$ incident edges and whose faces are colored in black for the lower index ranging from 1 to $N$ and white for the upper index ranging from 1 to $R$, so that two neighboring faces have different colors \cite{Anderson:1991ku, Francesco-rect}. If the black and white faces are respectively counted by $F_b$ and $F_w$, and assuming that $R\sim N^\alpha$ with $\alpha\ge 1$ (for $0<\alpha<1$ the roles of $N$ and $R$ are just exchanged),  the dependence in $\lambda$, $N$, $R$ of a graph behaves as $$\lambda^n N^{F_b + \alpha F_w} = \lambda^n N^{\alpha(F_b + F_w) + (1-\alpha) F_b} = \lambda^nN^{\alpha(2+ n(p-1) -2g ) + (1-\alpha ) F_b},$$ 
  so that  we obtain the two following cases:
  \begin{enumerate}[label=--]
  \item If $\alpha = 1$, the dominant graphs are all the planar $2p$-regular ribbon graphs, and we recover the 2D quantum gravity phase \cite{2DgravityReview},
  \item If $\alpha \neq 1$, the dominant graphs are all the planar $2p$-regular ribbon graphs which in addition have a single black face (or a single white face if $\alpha < 1$). Such graphs are easily shown to have the same structure as the dominant graphs for a $(\phi.\phi)^p$ vector model, which we describe in  Appendix~\ref{app:proof} for the $(\phi.\phi)^3$ model. These graphs have a tree-like structure characteristic of the branched polymer phase.
  \end{enumerate}
  Note that by scaling the coupling constant as $\lambda= \lambda' N^{\alpha(1-p)}$, the contributions of the graphs are bounded by $N^{1+\alpha}$.
The scenario for dominant graphs for the rectangular one-matrix model with the assumption $R\sim N^\alpha$ with $\alpha> 0$ is summarized in Fig.~\ref{fig:RecDomMat}.

\begin{figure}[h!]
 \begin{center}
 \includegraphics[scale=0.8]{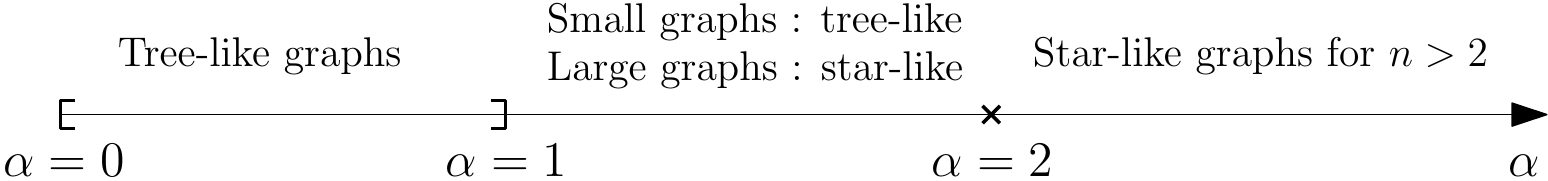}
 \caption{Dominant graphs at large $N$ for our random $N\times N^\alpha$ matrix models with non-pairwise index contractions with $\alpha> 0$.}
 \label{fig:RecDom}
 \end{center}
 \end{figure}

In this section, we will show that the scenario for the dominant graphs of our model is as shown in Fig.~\ref{fig:RecDom}. The families of graphs referred to as tree-like and star-like will be described more precisely in the rest of the section.    
 
We will see that while the dominant graphs for finite $R$ and large $N$ are the same as those for $R\sim N^\alpha$ with $0<\alpha\le 1$, the dominant graphs for $R\sim N^\alpha$ with $\alpha> 2$ are a strict subset of those for $R$ large and $N$ finite. In the intermediate regime where $R\sim N^\alpha$ with $1<\alpha\le 2$, there is a competition between the two families of dominant graphs, so that dominant graphs are neither included in those for $R$ large and $N$ finite, nor in those for finite $R$ and large $N$.

\subsection{The large $N$ and finite $R$ regime}
\label{sub:LargeNfiniteR}

\noindent{\bf The $R=1$ vector model at large $N$.} Let us start with this well-known particular regime of the model: for $R=1$,  we recover the $(\phi\cdot \phi)^3$ vector model, for which the graphs that maximize $\Fd$ in $\bG_c(n)$ are well-known and satisfy $\Fd=1+2n$ (an example is shown in Fig.~\ref{fig:Tree}). Such graphs are said to have a tree-like structure (see Appendix~\ref{app:proof}).

\begin{figure}[h!]
 \begin{center}
 \includegraphics[scale=0.5]{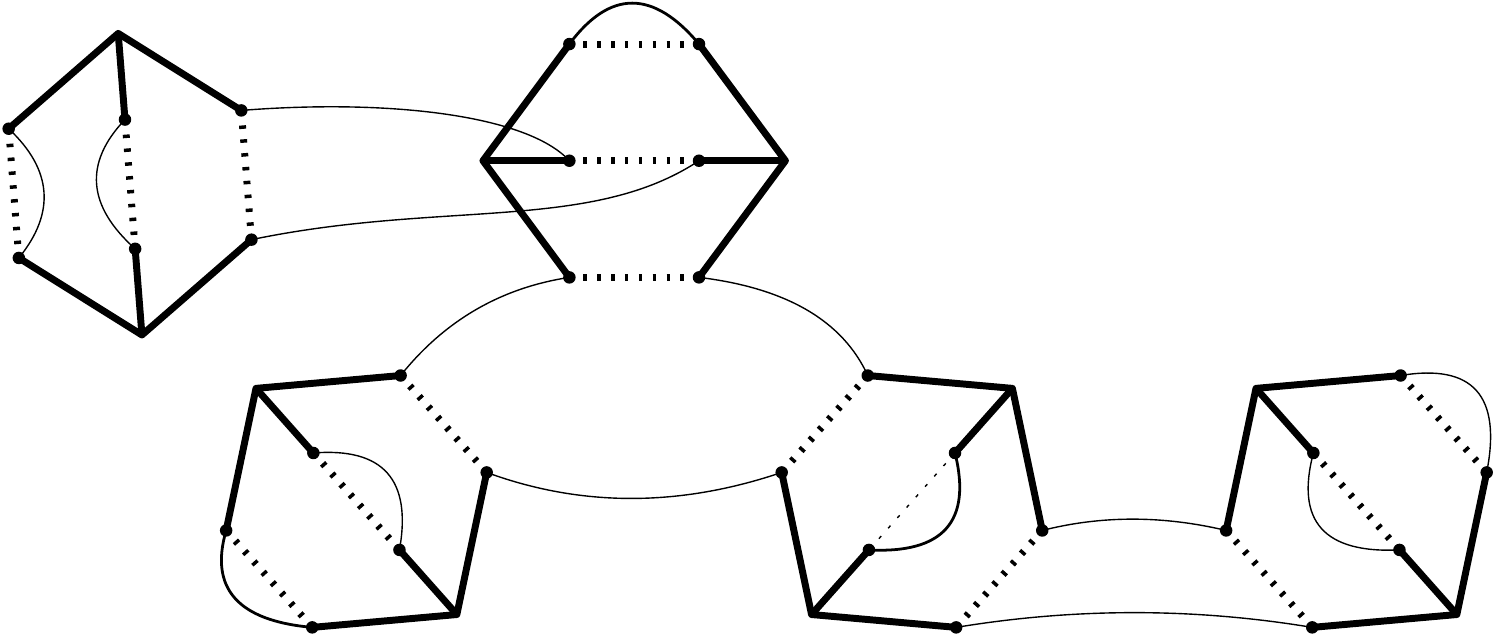}
 \caption{An example of dominant graph in the large $N$ and finite $R$ regime.}
 \label{fig:Tree}
 \end{center}
 \end{figure}

To compute the free-energy restricted to the dominant graphs in this regime, it is easier to first compute the 2-point function. More precisely, by differentiating the free-energy with respect to $k$, we obtain the generating series  $ \cG_{N,R}(\lambda, k) $ of graphs 
with one oriented thin edge, which corresponds to the \textit{normalized two-point function}
\be 
 \label{eq:TwoPoint-diff}
 \cG_{N,R}(\lambda, k)= \frac{2k}{NR}\langle \Tr \phi \phi^t \rangle  = 1- \frac{2k}{NR} \frac{\partial}{\partial k} \cF_{N,R}(\lambda, k)  ,
  \ee
  where 
  \be 
  \langle \Tr \phi \phi^t \rangle= \frac1{ \cZ_{N,R}(\lambda, k) } \int_{\bR^{NR}}\, d\phi \Tr \phi \phi^t  e^{-\lambda U(\phi) - 
k \Tr (\phi \phi^t)}.
  \ee

 \begin{figure}[h!]
 \begin{center}
 \includegraphics[scale=0.55]{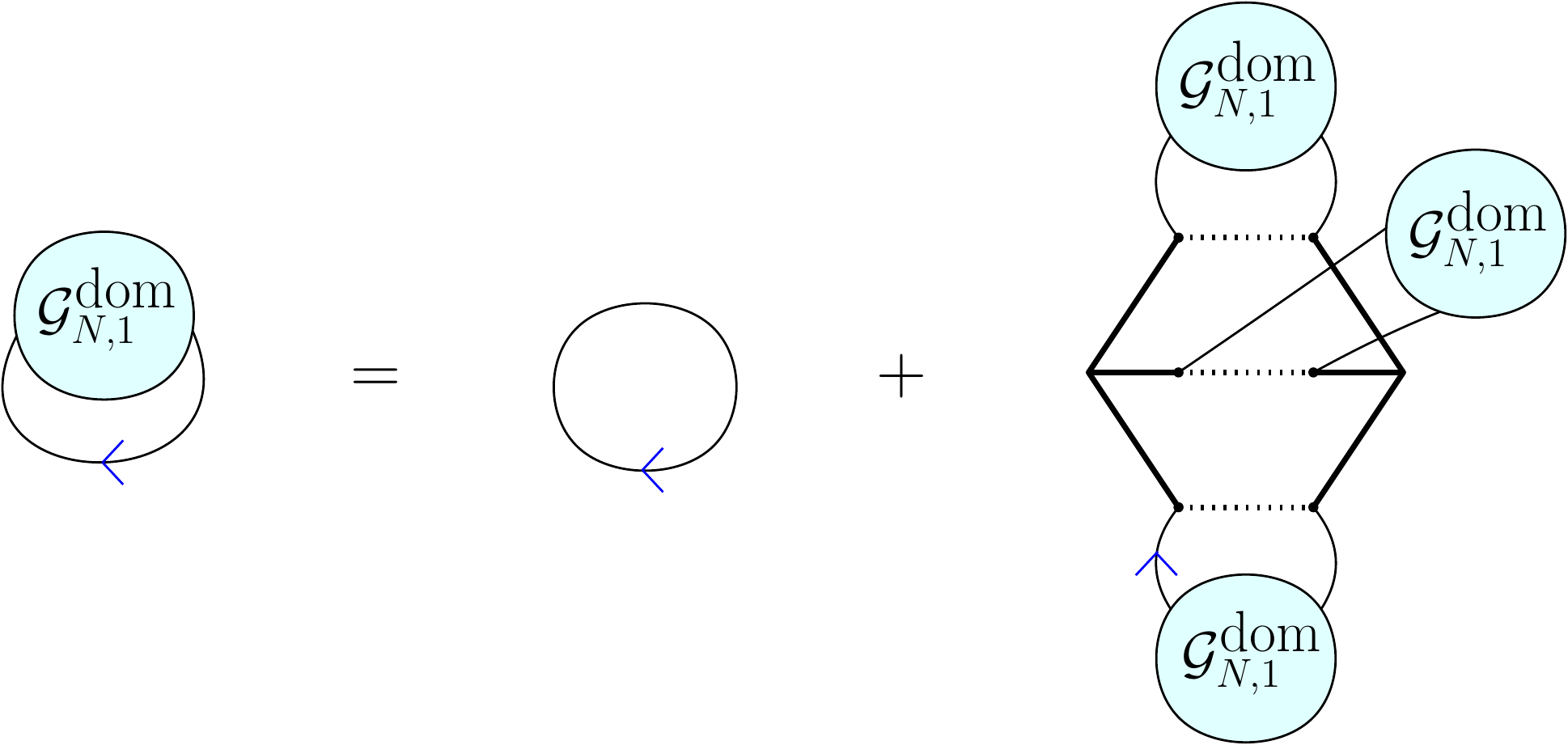}
 \caption{Graphical representation of the self-consistency equation for the normalized two-point function at leading order in the large $N$ and $R=1$ regime.}
 \label{fig:Two-point-trees}
 \end{center}
 \end{figure}

The dominant graphs in this regime have the recursive structure shown in Fig.~\ref{fig:Two-point-trees}, which translates in the following self-consistency equation for the 2-point function $\cG^\dom_{N,1}$ restricted to dominant graphs,
\be 
\label{eq:Ternary-Tree}
 \cG^\dom_{N,1} =  1 + z \times (\cG^\dom_{N,1})^3, \qquad z= -\frac {3\lambda N^2}{4k^3},
\ee 
where the factor $N^2$ comes from the normalization $1/NR$ in \eqref{eq:TwoPoint-diff} (see Appendix~\ref{app:proof}). This is the usual self-consistency equation for the generating function of rooted regular ternary trees. Note that by choosing the dependence in $N$ $\lambda=N\lambda'$ and $k=N k'$ where $\lambda', k'$ do not depend on $N$,  as usually done for vector models, we see that $ \cG^\dom_{N,1}$ no longer depends on $N$, so that  we get a well-defined limit when $N$ goes to infinity.

The coefficients of $ \cG^\dom_{N,1} (\lambda, k)$  are obtained using Lagrange inversion, and are known to be Fuss-Catalan numbers
\be 
 \cG^\dom_{N,1} (\lambda, k) = \sum_{n\ge 0} \frac 1 {3n + 1}\binom{3n+1}{n}\Bigl(-\frac {3\lambda N^2}{4k^3}\Bigr)^n.
\ee 
 
By integrating over $k$, we find the coefficients of the leading-order free energy,
\be
\label{eq:series-lo-FE}
 \cF^\dom_{N,1}(\lambda, k) =  \frac{N}6\sum_{n\ge 1} \frac 1 {n }\frac 1 {3n + 1}\binom{3n+1}{n}\Bigl(-\frac {3\lambda N^2}{4k^3}\Bigr)^n,
\ee
where the integration constant has been determined from the fact that $ \cF^\dom_{N,1}(0, k)=0 $.

The equation \eqref{eq:Ternary-Tree} can also be solved explicitly, and the solution to this equation for $z<0$ which 
leads to the right series expansion is 
\be
\label{eq:solTernary}
 \cG^\dom_{N,1} (\lambda, k) = -\frac{2\times 3^{1/3} z + 
  2^{1/3} (9 z^2 + \sqrt{3} \sqrt{z^3 (-4 + 27 z)})^{2/3}}{
 6^{2/3} z (9 z^2 + \sqrt{3} \sqrt{z^3 (-4 + 27 z)})^{1/3}) }.
\ee 

To recover an exact expression for $ \cF^\dom_{N,1}$, one can use \eqref{eq:TwoPoint-diff} and integrate \eqref{eq:solTernary} over $k$, however this becomes quite cumbersome.  Rather, it is easy to express $ \cF^\dom_{N,1}$ as a function of $ \cG^\dom_{N,1}$ by keeping the latter to implicitly represent the $k$-dependence. We obtain 
\be
 \cF^\dom_{N,1} =- \frac N 2\Bigl(  \cG^\dom_{N,1} + \frac{\lambda N^2}{4k^3}   (\cG^\dom_{N,1})^3 - \log( \cG^\dom_{N,1}) - 1\Bigr),
\ee
where the integration constant is found knowing that $\cG^\dom_{N,1}(0,k)=1$ and $ \cF^\dom_{N,1}(0, k)=0 $. One can easily show $ \cF^\dom_{N,1}$ satisfies \eqref{eq:TwoPoint-diff} due to \eqref{eq:Ternary-Tree}.

\

\noindent{\bf The finite $R$ case at large $N$.} In this case, the dominant graphs are the same as the $R=1$ case, the only difference being that we need to take into account the factor $R^{\Fs}$ in \eqref{eq:free-nrj}. Because of the tree-like structure (Figs.~\ref{fig:Tree} and \ref{fig:Two-point-trees}), it is easily seen that the graphs that maximize $\Fd$ in $\bG_c(n)$ have $\Fs = 1$, so that for $R$ finite, at large $N$, $\cG^\dom_{N,R}(\lambda, k) = \cG^\dom_{N,1}(\lambda, k)$ and $\cF^\dom_{N,R}(\lambda, k) = R \cF^\dom_{N,1}(\lambda, k)$.

\

A consequence is that considering random coupling vector models with a finite number $R$ of real replicas of the form
\[
\label{eq:random-coup-2}
Z_{N,R}(\lambda, k) = \int dP e^{-\frac 1 2 \sum_{abc=1}^N P_{abc}^2}\Bigl(\int_{\bR^N} \prod_a d\phi_a e^{ - k \sum_{a=1}^N \phi_a^2  - I\sqrt{2\lambda } \sum_{a,b,c=1}^N P_{abc} \phi_a \phi_b \phi_c }\Bigr)^R,
\] 
introducing replicas of the fields $\phi^i$, $(i=1,\ldots,R)$, and then expanding over Feynman graph, the graphs that dominate at large $N$ are the celebrated melonic graphs \cite{melons1, melons2, melons3}. This is explained in more details in Appendix \ref{app:melo}.

\subsection{The large $R$ and finite $N$ regime}
\label{sub:LargeRfiniteN}

\subsubsection{Results}

In the case where $R$ is large and $N$ is kept finite, we want to identify the graphs which maximize $\Fs$ in $\bG_c(n)$. We will show that in this regime,  the sum of the contributions of the dominant connected Feynman graphs in $\bG_c(n)$  for any $n\ge 1$ is given by 
\be
 \cF^\dom_{N,R}(\lambda, k) = \sum_{n\ge 1}\biggl[\frac{N}{2n}\Bigl(-\frac { 6(N+4) \lambda R}{8k^3}\Bigr)^n+\frac{N^3+3 N^2-4N}{12 n}\Bigl(-\frac {12 \lambda  R}{8k^3}\Bigr)^n \biggr].
 \label{eq:dnr}
\ee
Summing this series, we find the dominant free-energy to be, in this regime,
\[ 
\cF^\dom_{N,R}(\lambda, k) &= 
- \frac N 2 \log\Bigl(1+\frac{3(N+4)R\lambda }{4k^3}\Bigr)
- \frac{N(N+4)(N-1)}{12} \log\Bigl(1+\frac{3R\lambda }{2k^3}\Bigr).
\label{eq:freeenergylargeR}
\] 
Note that we can choose the dependence in $R$ of $\lambda$ and $k$ in order to cancel the dependence in $R$ and have a well defined limit for $R\rightarrow \infty$ and $N$ finite, e.g. by choosing  $\lambda=\lambda'/R$ with  $\lambda', k$ independent of $R$, and $\lvert \lambda'/k^3\rvert < 4/(3(N+4))$

By exponentiation, we find the dominant partition function in this regime to be  
\[
 \cZ^\dom_{N,R}(\lambda, k)= 
 \left( 1+ \frac{3(N+4)R\lambda}{4 k^3}\right)^{-\frac{N}{2}}
\left(
1+ \frac{3R\lambda}{2k^3}
\right)^{-\frac{N(N+4)(N-1)}{12}}.
\label{eq:resultsum}
\]

\begin{figure}[!h]
 \begin{center}
 \includegraphics[width=3cm]{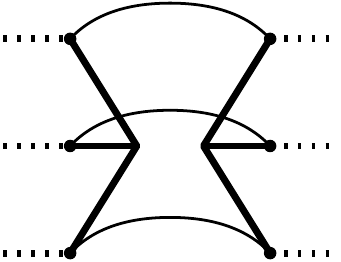}
 \hfil
 \includegraphics[width=3cm]{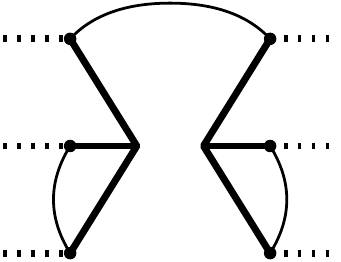}
 \caption{The building blocks of the dominant graphs at large $R$ and finite $N$. These building blocks can be obtained by cutting the dotted edges in half in the graphs of Figure~\ref{fig:neq1}.
In a graph made of these building blocks, the three dotted half-edges on each side must be paired with those
 on another building block, or itself.}
 \label{fig:components}
 \end{center}
 \end{figure}

{\noindent \it Dominant graphs. }In the finite  $R$ case at large $N$, it was possible to have dotted faces with a single dotted edge. This gives rise to the tree-like structure of the dominant graphs. In the present case however, a solid face necessarily has an {\it even} number of trivalent solid nodes. Furthermore, $\Fs$ is bounded from above by the number of interactions
\be
\text{if }G\in\bG_c(n), \qquad \Fs(G) \le n(G), 
\ee
with equality if and only if every solid face has exactly two trivalent solid nodes. The dominant graphs in the large $R$ and finite $N$ regime are thus the graphs which satisfy this last condition, and they are easily shown to be necklace-like graphs obtained by forming one loop with the building blocks listed in Figure~\ref{fig:components} (see the examples in Figure~\ref{fig:necklace}).

\subsubsection{Proof}

\begin{figure}[!h]
 \begin{center}
 \includegraphics[width=5cm]{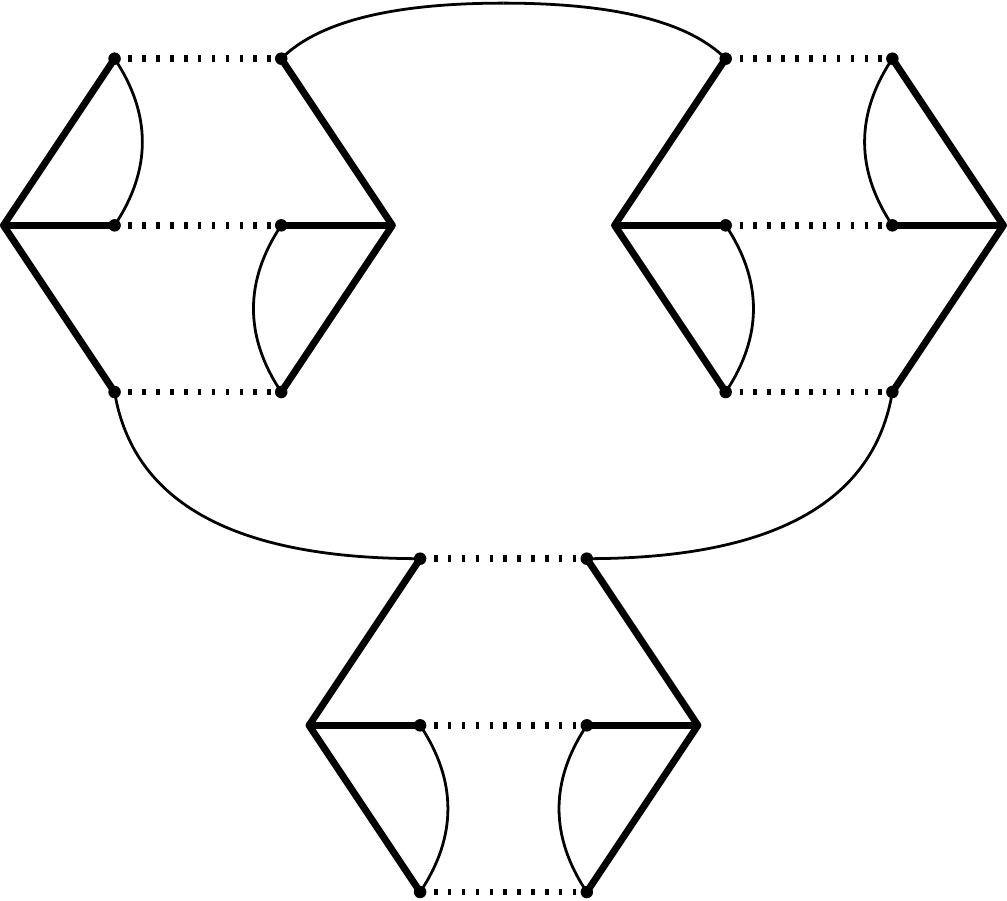}
 \hfil
 \includegraphics[width=5cm]{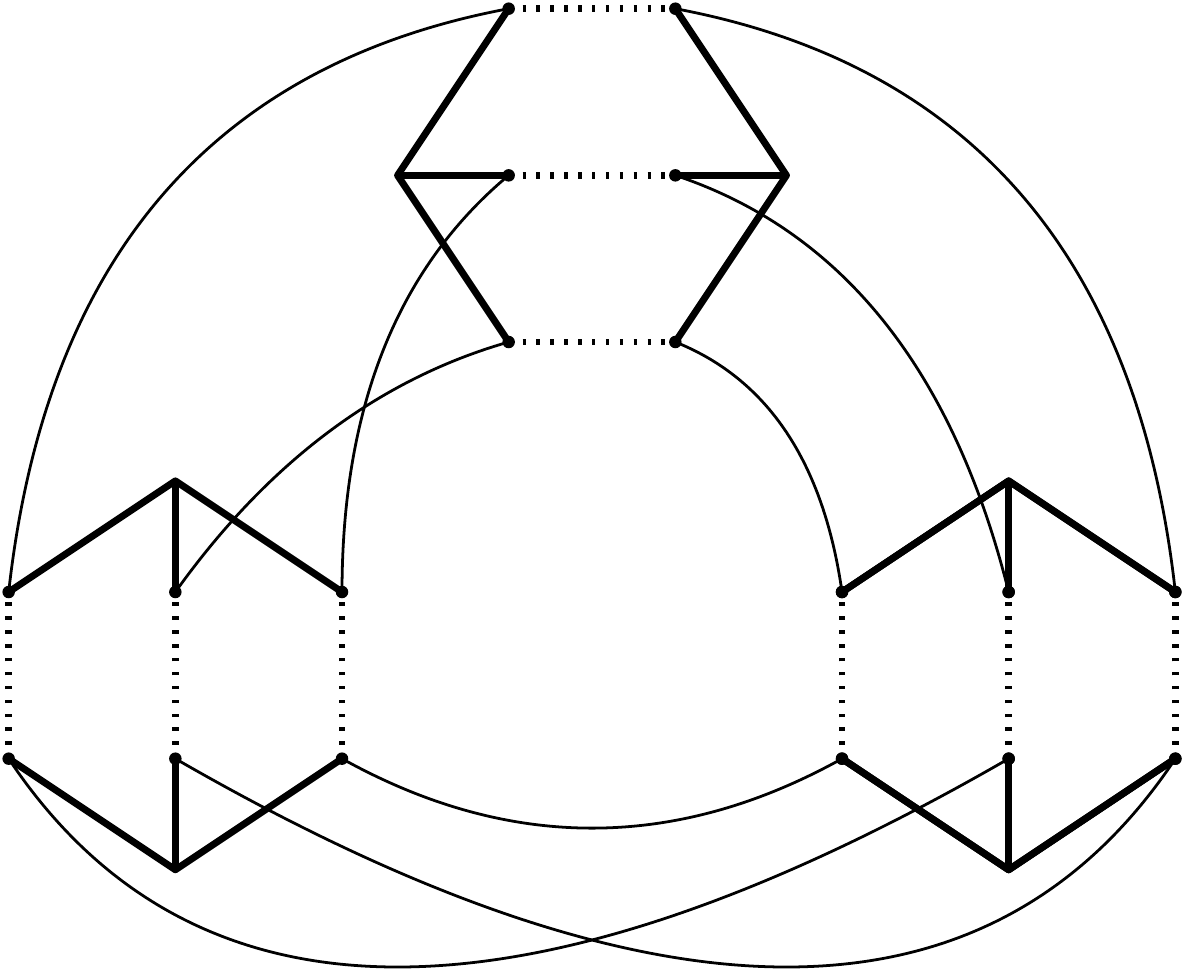}
 \caption{Examples of necklace graphs for $n=3$. 
 The weights in $N,R$ are $N^2 R^3$ and $NR^3$, respectively. }
 \label{fig:necklace}
 \end{center}
 \end{figure}

More precisely, given a graph in $G\in\bG_c(n)$, let us consider the following abstract graph $\Gamma(G)$: for each solid face $f$ we draw a vertex $v(f)$, and for each interaction in $G$, if its two 3-valent solid nodes belong to some (non-necessarily distinct) faces $f_1$ and $f_2$, we draw an edge between the corresponding vertices $v(f_1)$ and $v(f_2)$. Then the number of independent loops in the  graph $\Gamma$ is $L(\Gamma) = n(G) - F_s(G) +1$, since $\Gamma$ has $n(G)$ edges, $F_s(G) $ vertices, and is connected. Therefore, 
\be 
F_s(G) = n (G) +1 - L(\Gamma). 
\ee
Furthermore, graphs with no loops are trees, which necessarily have vertices of valency one.  But as the solid faces in $G$ contain an even number of 3-valent solid nodes, the vertices in $\Gamma(G)$ have at least valency two. Therefore, $L(\Gamma)\ge 1$, and we recover that  $1\le \Fs(G) \le n(G)$, but in addition we now know that the graphs in $\bG_c(n)$ whose contribution in $R$ is $R^{n+1-l}$ are obtained by considering all the abstract graphs $\Gamma(G)$ with $l$ loops. In theory, we can thus identify the graphs contributing at any order in $R$. In particular, as written above, the dominant contribution in $R^n$ is given by the graphs for which $\Gamma$ is the only one-loop graph, which corresponds to the necklaces  of  building blocks listed in Fig.~\ref{fig:components}.  Two examples of necklace graphs for $n=3$ are shown in Fig.~\ref{fig:necklace}.
\vspace{-0.5cm}
\vspace{-0.5cm}

As was just proven, the dominant graphs are such that the solid faces having exactly two trivalent solid nodes are connected by  dotted edges to form a loop. Such solid faces are the building blocks of the dominant graphs, and  there exist only two kinds shown in 
Fig.~\ref{fig:components}.
These building blocks can be obtained by cutting the dotted edges in half in the graphs of Fig.~\ref{fig:neq1}.
Two building blocks are connected  by the  dotted half-edges  on one of their sides (or both, if the whole graph is composed of a single building block). The summation over Wick pairings in such a building block can be accounted by  permuting the dotted half-edges.
To count the number of ways of connecting the building blocks in a loop, it is convenient to use a matrix representation. By this, the 
free energy coming from the dominant  graphs with 
$n$ interactions is given by 
\[
f^\dom_n=\frac{2^n(-\lambda)^n R^n}{2 n(2k)^{3n}} {\rm Tr}\left( A^n\right),
\label{eq:fn}
\]
where $A$ is a matrix representing the connection of the dotted edges  of a building block. More precisely, the matrix $A$ is a sum of two matrices, $A=B+C$, respectively corresponding to the two kinds of building blocks in Figure~\ref{fig:components}:
\[
\begin{split}
&B:=\frac{1}{6}P\tilde B P,\ \  \tilde B_{abc,def}:=\delta_{ad}\delta_{be}\delta_{cf},\\
&C:=\frac{1}{4}P\tilde CP, \ \ \tilde C_{abc,def}:=\delta_{ad} \delta_{bc}\delta_{ef},
\end{split}
\label{eq:ABC}
\]
where a product of two matrices, say $X$ and $Y$,
is defined by
$$(XY)_{abc,ghi}:=\sum_{d,e,f=1}^N X_{abc,def}Y_{def,ghi},$$ 
and 
\be
P_{abc,def}:=\delta_{ad}\delta_{be}\delta_{cf}
+(\hbox{permutations of }d,e,f).
\ee
The matrix $P$ represents all permutations of the three dotted edges on each side of a building block. The numerical factors in \eq{eq:ABC} cancel the graph degeneracies\footnote{Among the 36 terms generated, some correspond to the same graphs. There are respectively 6 and 9 non-equivalent terms for $B$ and $C$, with degeneracies 6 and 4.}.

As for the other factors in \eqref{eq:fn}, the factor $2^n$ comes from the choice of the sides of the interactions to form the solid faces, the factor $R^n$ accounts for the contribution of the $n$ solid faces, and there is a symmetric factor $2n$ in the denominator, where 2 comes from the overall reflection and $n$ from the choice of starting points in a loop.

To compute  \eq{eq:fn}, we use the properties of $B,C$ and $P$.
By using $P^2=6 P,\ P\tilde B=\tilde BP=P$ (so that $B=P$), and $\tilde CP\tilde C=2(N+2)\tilde C$, 
one obtains 
\be
B^2=6B,
\quad BC=CB=6C,\quad C^2=3(N+2)C.
\label{eq:bcprod}
\ee
One can also show
\be
{\rm Tr}(B)=N^3+3N^2+2N,\quad 
{\rm Tr}(C)=3N(N+2).
\label{eq:bctr}
\ee
Though $B$ and $C$ are the natural choices 
for representing the connections of the dotted edges, they are not convenient
for the computation of \eq{eq:fn}  because of the mixed structure of their products.
A better choice is given by
\[
K=\frac{1}{6}B-\frac{1}{3(N+2)}C,\ \  H=\frac{1}{3(N+2)}C.
\]
Indeed, from \eq{eq:bcprod} and \eq{eq:bctr},
these quantities satisfy
\[
\begin{split}
& K^2=K,\qquad HK=KH=0,\qquad  H^2=H, \\
& \hbox{Tr}(K)=\frac{1}{6}(N^3+3N^2-4N),\quad  \hbox{Tr}(H)=N.
\end{split}
\]
Since $A=6 K +3 (N+4) H$, we obtain
\[
{\rm Tr}(A^n)=6^{n-1}(N^3+3N^2-4N)+3^n(N+4)^n N. 
\]
By putting this into \eq{eq:fn}, we obtain the forementioned result \eq{eq:dnr}.

\

\begin{figure}[h!]
 \begin{center}
 \includegraphics[scale=0.4]{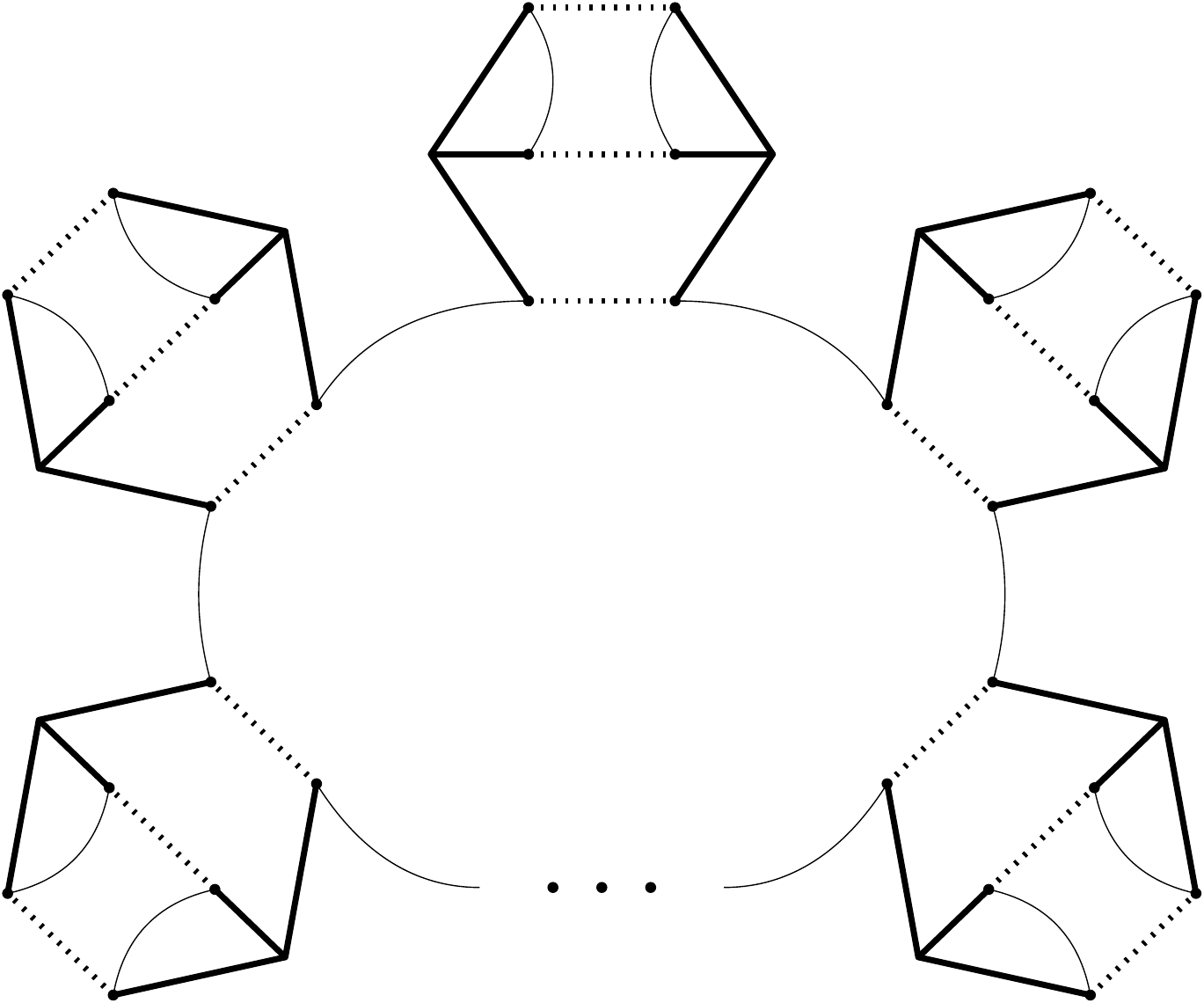}
 \caption{ The star-like graphs which dominate for sufficiently large $n$ in the large $R\sim N^\alpha$  regime with $\alpha>1$.}
 \label{fig:LO-largNR}
 \end{center}
 \end{figure}

Note that the graphs which maximize $\Fd$ at fixed $n$, when $\Fs=n$ satisfy $\Fd=n+1$. If there are only building blocks as on the left of Fig.~\ref{fig:components}, the number of dotted faces is bounded by 3. To obtain $\Fs=n+1$ for $n>2$, we therefore need to have building blocks as on the right of Fig.~\ref{fig:components}, which means that there is a single dotted face going around the loop. The number of the remaining dotted faces is bounded by $n$, which occurs when all the building blocks are as on the right of Fig.~\ref{fig:components}, and the dotted edges produce one dotted face between every two building blocks. This imposes the graph to be as in Fig.~\ref{fig:LO-largNR}. We will call such graphs {\it star-like graphs}.

\subsection{The large $R\sim N^\alpha$  regime with $\alpha>1$}
\label{sub:AlphaLarg1}

\subsubsection{Results}

In this section, we will identify the dominant graphs in the regime where $R\sim N^\alpha$  with $\alpha>1$ and $N\rightarrow +\infty$. They correspond to the graphs in $\bG_c(n)$ which maximize 
\be
F_\alpha = \Fd + \alpha \Fs,
\ee
 with $\alpha>1$. We have seen two families of graphs which maximize $\Fd$ at fixed $\Fs$:
 \begin{itemize}
 \item The \emph{tree-like graphs}, which have a maximal $\Fd$ among all graphs at fixed $n$, and for which $\Fs=1$. Tree-like graphs in $\bG_c(n)$ thus have 
 \be
 F_{\alpha,n}^\text{tree} = 1+\alpha + 2n
 \ee
 \item The \emph{star-like graphs} shown in Fig.~\ref{fig:LO-largNR}, which have a maximal $\Fd$ at fixed $n$ ($\Fd=n+1$), among graphs with maximal $\Fs$. Star-like graphs in $\bG_c(n)$ thus have 
 \be
 F_{\alpha,n}^\text{star} = 1+(\alpha + 1)n.
 \ee
 \end{itemize}
 There is a competition between the two families of graphs. Indeed, we see that
 \be 
 \label{eq:Comp-star-tree}
  F_{\alpha,n}^\text{tree} \le F_{\alpha,n}^\text{star} \qquad \Leftrightarrow \qquad n\ge \frac \alpha{\alpha - 1}.
 \ee

\begin{figure}[h!]
 \begin{center}
 \includegraphics[scale=0.8]{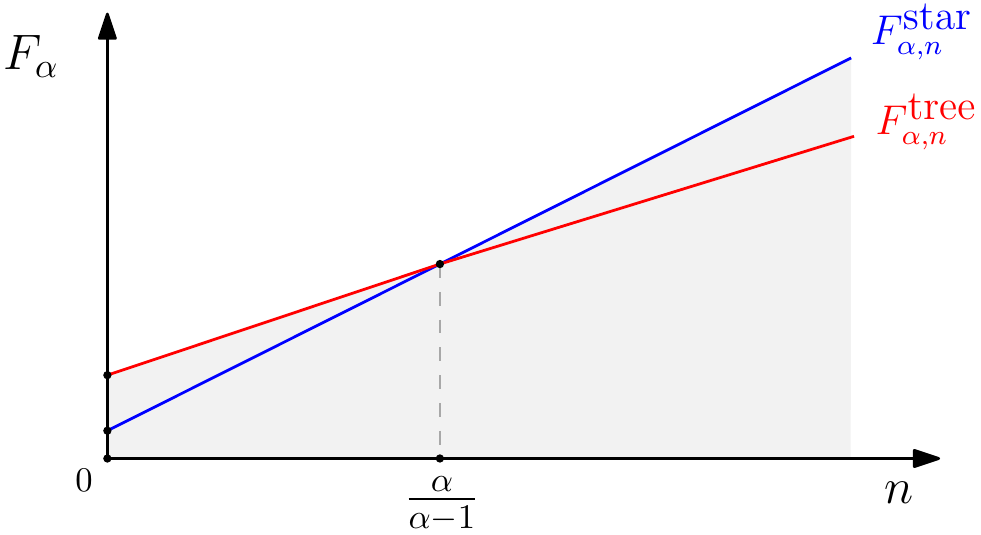}
 \caption{$F_\alpha$ as a function of $n$ for $\alpha>1$. The region of reachable $F_\alpha$ is shaded. It is delimited by  $ F_{\alpha,n}^\text{tree}$ for small $n$ and by  $F_{\alpha,n}^\text{star}$ for larger $n$.}
 \label{fig:fig-Falpha}
 \end{center}
 \end{figure}

 In this section, we will show that all other graphs are dominated either by the tree-like graphs or by the star-like graphs, in the sense that they have a lower $F_\alpha$ at fixed $n$. An exception occurs for $n=2$, for which another one of the necklace graphs has the same contribution in $N$ and $R$ as the star-like graph (it belongs to the family of necklaces whose contribution is in $N^3R^n$ in \eqref{eq:dnr}).
Therefore, \eqref{eq:Comp-star-tree} describes the unusual scenario for dominant graphs, which we summarized in Fig.~\ref{fig:RecDom} and Fig.~\ref{fig:fig-Falpha}: 
\begin{enumerate}[label={\small $\blacktriangleright$}]
\item \underline{For $\alpha>2$}, we have $1<\alpha/(\alpha - 1) <2$, so that tree-like graphs only dominate at $n=1$, while star-like graphs dominate for $n> 2$. Both the star-like graphs and the necklaces whose contribution is in $N^3R^n$ in \eqref{eq:dnr} co-dominate at $n=2$. As a consequence, in this regime, the dominant free-energy  is given by 
\be
\cF^\dom_{N,R}(\lambda, k) = 
- \frac {N^2R\lambda}{ 8k^3}(N-3) + \frac{3 N^3 R^2\lambda^2}{32 k^6}- \frac N 2 \log\Bigl(1+\frac{3NR\lambda }{4k^3}\Bigr).
\label{eq:allarger2}
\ee
 \item \underline{For $\alpha=2$}, we have $\alpha/(\alpha - 1) =2$, so that tree-like graphs dominate at $n=1$, tree-like graphs, star-like graphs, and the necklaces whose contribution is in $N^3R^n$ in \eqref{eq:dnr} co-dominate at $n=2$, while star-like graphs  dominate for $n>2$. As a consequence, in this regime, the dominant free-energy  is given by 
\be
\cF^\dom_{N,R}(\lambda, k) = 
- \frac {N^2R\lambda}{ 8k^3}(N-3)  + \frac{3 N^3 R\lambda^2}{32 k^6}\bigl(\frac {3} 2 N^2 + R\bigr) - \frac N 2 \log\Bigl(1+\frac{3NR\lambda }{4k^3}\Bigr).
\label{eq:aleq2}
\ee
 \item \underline{For $1<\alpha<2$}, we have $\alpha/(\alpha - 1) >2$. For $n\le \frac \alpha{\alpha - 1}$, tree-like graphs dominate, while for $n\ge \frac \alpha{\alpha - 1}$, star-like graphs  dominate. If $\alpha=\frac {n_0}{n_0-1}$ for some positive integer $n_0$, then both tree-like graphs and star-like graphs  co-dominate for $n=n_0$. As a consequence,  in this regime, the dominant free-energy  is given by 
\be
\cF^\dom_{N,R}(\lambda, k) =\frac {NR} 6 \sum_{n=1}^{\lfloor \frac \alpha{\alpha - 1}\rfloor} \frac 1 {n }\frac 1 {3n + 1}\binom{3n+1}{n}\Bigl(-\frac {3\lambda N^2}{4k^3}\Bigr)^n + \frac N 2 \sum_{n\ge\lceil \frac \alpha{\alpha - 1}\rceil} \frac 1 n \Bigl(-\frac {3\lambda NR}{4k^3}\Bigr)^n.
\label{eq:a-bet-12}
\ee
 \end{enumerate}
 
\subsubsection{Discussion}
\label{sec:discussionlimit}

The series we obtain for the star-like graphs is the remainder of a logarithm, which is convergent if $\lvert \frac \lambda {k^3}\rvert < \frac 4 3 \frac 1 {NR}$ and divergent otherwise. By choosing the dependence in $N, R$ of the coupling constants to compensate the factors $(NR)^n$ in the sum corresponding to the logarithm, i.e.
\be 
\frac\lambda{k^3} = \frac{\lambda'}{k'^3} \frac 1 {NR}\quad\text{ with }\quad\Bigl\lvert\frac{\lambda'}{k'^3}\Bigr\rvert < \frac 4 3,
\label{eq:scaling-lamda-k}
\ee
the sum on the right of \eqref{eq:a-bet-12} can be replaced by the remainder of the logarithm, which scales in $N$:
\be
\label{eq:log-free-NRJ-1}
\cF^\dom_{N,R}\Bigl(\frac{\lambda'}{NR}, k' \Bigr) =
\frac {NR} 6 \sum_{n=1}^{\lfloor \frac \alpha{\alpha - 1}\rfloor}u_n\Bigl(-\frac {3\lambda' N}{4k'^3R}\Bigr)^n 
- \frac N 2 \sum_{n=1}^{\lceil \frac \alpha{\alpha - 1}\rceil - 1}
\frac 1 n \Bigl(-\frac {3\lambda' }{4k'^3}\Bigr)^n -  \frac N 2 \log\Bigl(1+\frac{3\lambda' }{4k'^3}\Bigr),
\ee
where $u_n= \frac 1 {n }\frac 1 {3n + 1}\binom{3n+1}{n}$.  The terms in the partial sum of the tree-like free energy behave in $N^{1+ \alpha -n(\alpha - 1) }$, so that these terms all have a stronger scaling in $N$ than the logarithm\footnote{Apart from the term for $\alpha/(\alpha - 1)$ if it is an integer.}. Note that the dominant free-energy as we defined it only retains the dominant graphs at fixed $n$, so that for $n\le \lfloor \frac \alpha{\alpha - 1}\rfloor$ there might be other graphs whose dependence in $N$ is stronger than $N$ with the choice \eqref{eq:scaling-lamda-k}.
However, there are only finitely many of them, so that there exists a polynomial 
  \be 
  \nonumber
  P_\alpha(\lambda'/k'^3, N, R) = \sum_{n=1}^{\lfloor \frac \alpha{\alpha - 1}\rfloor} c_n(N,R)\Bigl(-\frac {\lambda' }{8k'^3}\Bigr)^n,
  \ee 
  where $c_n(N,R)$ gathers the contributions of all the graphs with $n$ interactions whose dependence in $N$ when $R=N^\alpha$ is stronger or equal to $N$, aside from the star-like graphs, so that  
  $$
  c_n(N,R) = u_n\frac {NR} 6 \Bigl(-\frac {3\lambda' N}{4k'^3R}\Bigr)^n + o
  \bigl(N^{2 -(n-1)(\alpha - 1)}\bigr),
  $$
 and in particular, 
 \be 
 P_\alpha(\lambda'/k'^3, N, R) = 
 -\frac {\lambda' N^2}{8k'^3}+ o(N^2),
 \ee
and such that 
 \[
\lim_{ \substack{{N\rightarrow +\infty,}\\{R\sim N^\alpha,\ \alpha >1}}
} \frac 1 {N}\, \Biggl[\log Z_{N,R} \Bigl(\frac{\lambda'}{NR}, k' \Bigr)  - P_\alpha(\lambda'/k'^3, N, R) \Biggr] =  -  \frac 1 2 \log\Bigl(1+\frac{3\lambda' }{4k'^3}\Bigr).
\label{eq:limitsub}
\]
We have for instance for any $\alpha>2$,\footnote{Note that because of the convention that $\cF^\dom_{N,R}$ only retains the dominant graphs order per order, we had to add the term of order 1 of the logarithm in \eq{eq:allarger2}, thus the  $-\frac {3\lambda' N^2}{8k'^3}$. This is not necessary here, as we deal with the full free-energy.} 
$$ P_{\alpha>2}(\lambda'/k'^3, N, R) =  -\frac {\lambda' N^2}{8k'^3} + \frac {3\lambda'^2N}{32k'^6},$$
and for $\alpha=2$, 
$$ P_{2}(\lambda'/k'^3, N, R) =  -\frac {\lambda' N^2}{8k'^3} +  \frac{3 N \lambda'^2}{32 k'^6}\Bigl(\frac {3} 2 \frac {N^2} R + 1\Bigr).$$ Said otherwise, after retrieving the contribution of a finite number of graphs, the large $N$ free energy is essentially a logarithm. In matrix models in the context of 2D quantum gravity, one is naturally interested in the behavior of large graphs, as they carry the properties of the continuum limit. Here, for large graphs, the free-energy is dominated by a logarithm, so that in a sense the large dominant graphs are more ordered than the smaller ones. This is an interesting phenomenon, which, as far as we know, had not been exhibited by any previously studied random vector, matrix or tensor model. 

\

Note that because the series of star-like graphs is highly divergent outside of its domain of convergence, the conclusions of the graphical study performed in this section (the identification of the various series of dominant graphs and the comparison between them) cannot be extrapolated outside the domain of convergence. This applies for instance to the regime where $\lambda/k^3 = t/N^2$ for $t$ of the order of 1. 

\subsubsection{Proof}

We split the proof in several parts.

\medskip

{\noindent \bf (a) \ A bound on the number of small faces. }We will need the following result for $n\ge 1$, which proves that {\it a dominant graph necessarily contains small faces}:
 \be 
 \label{eq:Bound-small-faces}
G \in \bG_c(n)\text{ is dominant}\quad \Rightarrow \quad\Fd[(1)](G) + \alpha \Fs[(2)](G) \ge 2 + n(\alpha - 1)
 \ee  
 where $\Fd[(l)]$ (resp.~$\Fs[(l)]$) is the number of dotted  (resp.~solid) faces incident to  $l$ interactions counted with multiplicity: for dotted faces, it means having $l$ dotted edges, and for solid faces, it means having $l$ trivalent solid nodes. 
 The proof of this preliminary result goes as follows.  In the following, we consider a graph $G\in\bG_c(n)$. By summing $l \Fd[(l)]$ (resp.~$l \Fs[(l)]$) over $l$, we just count the total number of dotted edges (resp.~trivalent solid nodes) in the graph:
\be
\label{eq:sum_lengths_faces}
 \sum_{l\ge 1}l \Fd[(l)](G) = 3n, \qquad\text{and}\qquad \sum_{l\ge 1}2l \Fs[(2l)](G) = 2n,
\ee
where the second sum has been restricted to even integer, due to the fact that solid faces visit an even number of interactions. By considering  $3n - \Fd(G)$, we find
$$
3n - \Fd= \sum_{l\ge 2} (l-1) \Fd[(l)] \ge \Fd - \Fd[(1)],
$$
so that 
\be 
\label{eq:bound-small-dotted}
\Fd[(1)] \ge 2\Fd - 3n.
\ee
with equality if and only if $\Fd[(l)]$ vanishes for $l>2$. 
Similarly, by considering  $n - \Fs(G)$, we find
$$
n - \Fs= \sum_{l\ge 2} (l-1) \Fs[(2l)] \ge \Fs - \Fs[(2)].
$$
so that 
\be 
\label{eq:bound-small-solid}
\Fs[(2)] \ge 2\Fs - n.
\ee
From \eqref{eq:bound-small-dotted} and \eqref{eq:bound-small-solid}, we see that any graph $G\in\bG_c(n)$ satisfies 
\be
\label{eq:small-bound-Falpha}
\Fd[(1)](G)  + \alpha \Fs[(2)](G) \ge 2F_\alpha(G) -  n(3 +   \alpha).
\ee

Since we know that the graph in Fig.~\ref{fig:LO-largNR} has  $F_\alpha = 1+(\alpha+1)n$, we know that $F_\alpha^\dom(n) \ge 1+(\alpha+1)n$, so that a dominant graph $G$ satisfies   
$$
\Fd[(1)](G)  + \alpha \Fs[(2)](G) \ge 2(1+(\alpha+1)n) -  n(3 +   \alpha),
$$
which simplifies to \eqref{eq:Bound-small-faces}.

\

We now prove recursively on the number of interactions $n$ that if $n<\frac \alpha {\alpha - 1}$,  the dominant graphs in $\bG_c(n)$ are the tree-like graphs of Sec.~\ref{sub:LargeNfiniteR}), and if $n>\frac \alpha {\alpha - 1}$,  the dominant graphs in $\bG_c(n)$ are the star-like graphs of Fig.~\ref{fig:LO-largNR}. The method is to assume that  $G$ is dominant, and to characterize it using some graphical moves. In the following, we assume that $n\ge4$, as we will initiate the induction at $n=3$. The cases $n=1,2,3$ will be treated below in the paragraph {\bf (d)}. Note that to show that a graph $G$  is a tree-like graph, it is sufficient to show that $\Fd(G)=2n(G)+1$, and to show that it is a star-like graph, if $n\ge 3$, it is sufficient to show that $\Fs(G)=n(G)$ and to assume that it is dominant.

\

{\noindent \bf (b) \  On the existence of  dotted faces with a single edge in a dominant graph. }
In this paragraph, we show that {\it if a graph $G$ contains a dotted face with a single dotted edge, then either 
$G$ is a tree, or $G$ is not dominant (i.e.~we can find another connected graph with a larger $F_\alpha$)}. 
Since tree-like graphs are not dominant for $n>\frac \alpha{\alpha - 1}$, this implies that:
\begin{enumerate}[label=(\roman*)]
\item\label{lemma11} {\it a dominant graph for $n>\frac \alpha{\alpha - 1}$ must satisfy $\Fd[(1)]=0$}.
\item\label{lemma12}{\it a graph with $n\le \frac \alpha{\alpha - 1}$ for which $\Fd[(1)]>0$ is  a tree-like graph or has a smaller $F_\alpha$ than tree-like graphs.}
\end{enumerate}

Suppose  that there exists a dotted face in $G$ with a single dotted edge. There are four possibilities locally, shown below.  
\begin{figure}[h!]
\center
\includegraphics[scale=0.45]{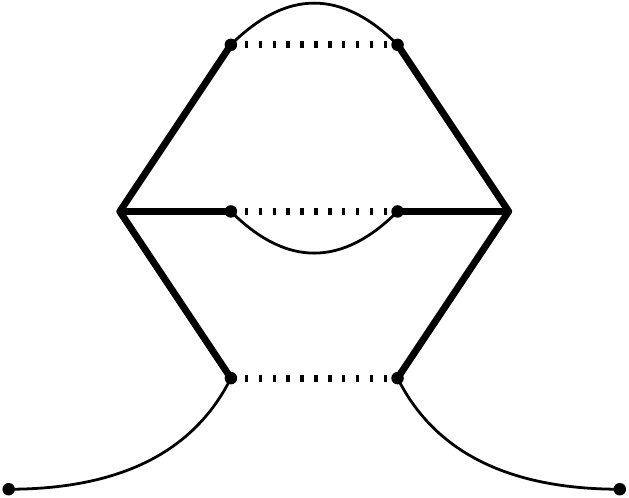}
\hspace{1cm}
\includegraphics[scale=0.45]{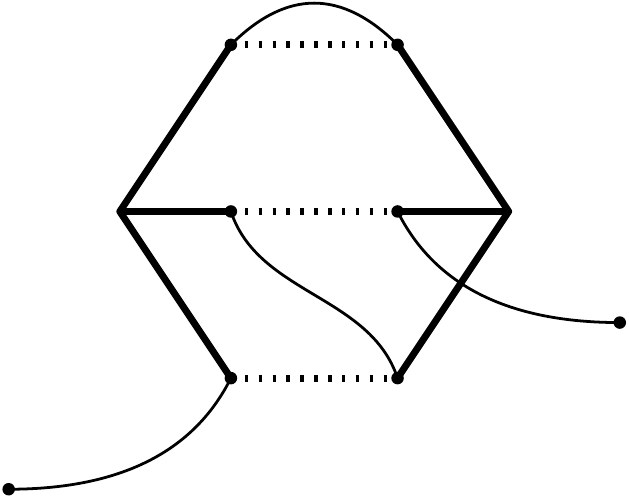}
\hspace{1cm}
\includegraphics[scale=0.45]{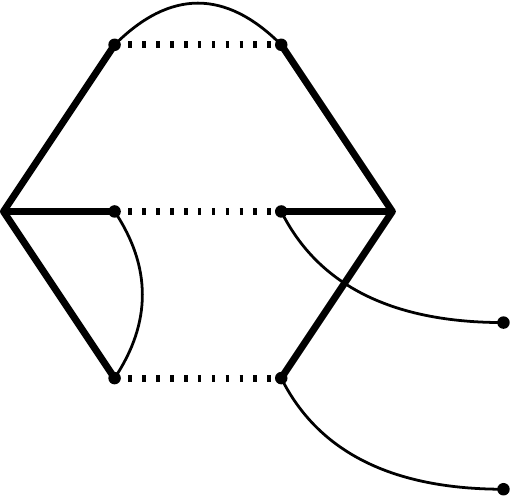}
\hspace{1cm}
\includegraphics[scale=0.45]{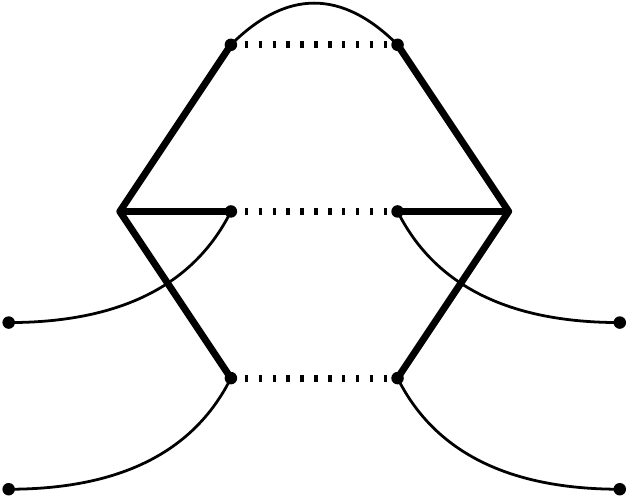}
\caption{Local possibilities for an interaction with a dotted face with a single dotted edge.}
\label{fig:One-dotted-cases}
\end{figure}

We immediately see that the case on the left has more dotted faces than the two cases in the middle, while they have the same number of solid faces. Therefore, a graph containing the subgraphs in the middle cannot be dominant. 
Let us first suppose that there exist a subgraph as on the left of Fig.~\ref{fig:One-dotted-cases}. Performing the operation below
\be
\label{fig:One-dotted-last-move}
\begin{array}{c}\includegraphics[scale=0.45]{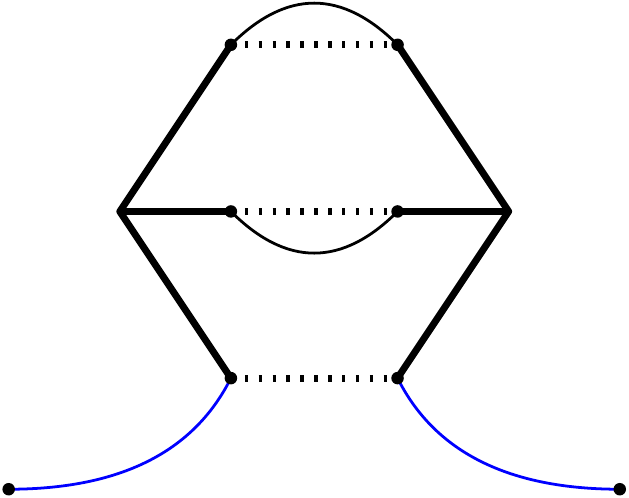}\end{array}
\hspace{1cm}\rightarrow\hspace{1cm}
\begin{array}{c}\includegraphics[scale=0.45]{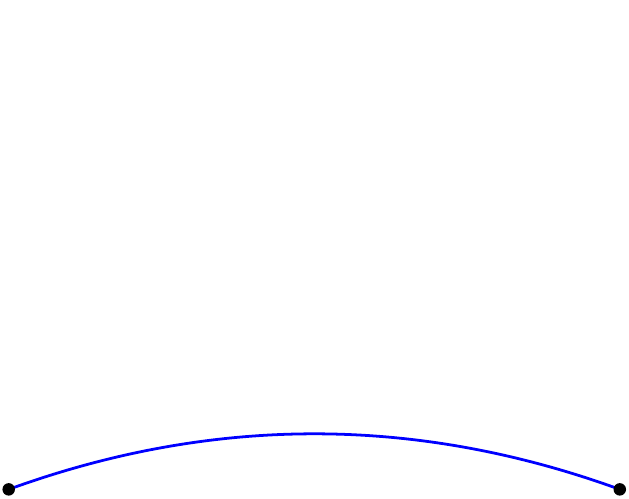}\end{array}
\ee 
we obtain a graph $G'$ with one less interaction. We have that $\Fd(G) = \Fd(G') + 2$ and $\Fs(G) = \Fs(G')$.

If $n-1< \alpha /({\alpha - 1})$, either $G'$ is a tree-like graph, in which case $G$ is a tree-like graph (and therefore $G$ is not dominant if $n>\frac \alpha {\alpha - 1}$), or using the recursion hypothesis, $F_\alpha(G')<1+\alpha + 2(n-1)$, so that  $F_\alpha(G) < 1+\alpha + 2n$, in which case $G$ is not dominant  (tree-like graphs always have a larger $F_\alpha$).

If $n-1\ge \frac \alpha {\alpha - 1}$, from the recursion hypothesis, $F_\alpha(G')\le1+(\alpha + 1)(n-1)$, so that $F_\alpha(G) \le 1+ (\alpha + 1)n  + 1 - \alpha < 1+ (\alpha + 1)n$, which implies that $G$ is not dominant (star-like graphs always have a larger $F_\alpha$).

Now focusing on the case on the right of Fig.~\ref{fig:One-dotted-cases}, we exchange the thin lines as illustrated below, for one of the two other dotted edges of the interaction:
\be
\label{fig:One-dotted-move}
\begin{array}{c}\includegraphics[scale=0.45]{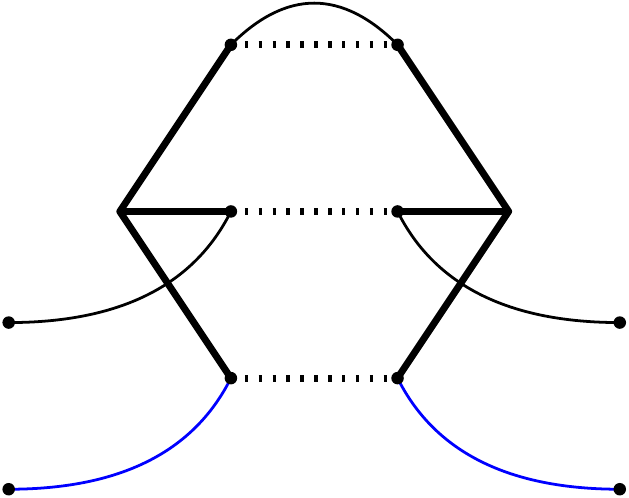}\end{array}
\hspace{1cm}\rightarrow\hspace{1cm}
\begin{array}{c}\includegraphics[scale=0.45]{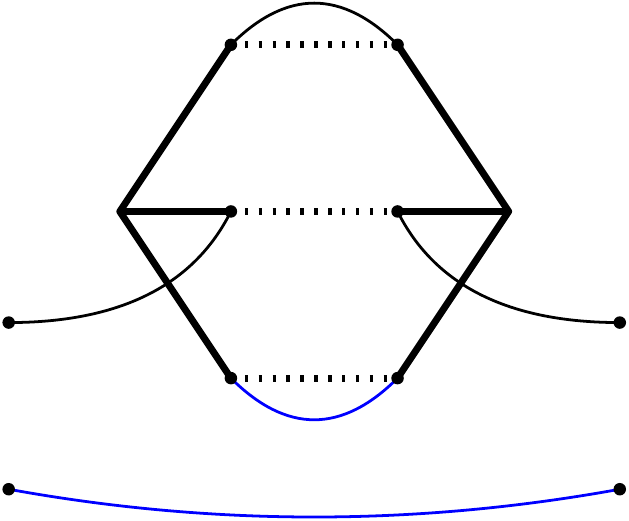}\end{array}.
\ee 
There are two cases. If this disconnects the graph into two graphs $G_1$ and $G_2$, one dotted face and one solid face are created, so that $F_\alpha(G) = F_\alpha(G_1) + F_\alpha(G_2)  - (1+\alpha)$. We bound $F_\alpha(G_1)$ and $ F_\alpha(G_2)$ by their maximal possible values, using the recursion hypothesis. Depending on whether $n_1=n(G_1)$ and  $n_2=n(G_2)$ are smaller than $\frac \alpha {\alpha - 1}$ or not, we may have the following situations. 
\begin{itemize}
\item If both  $n_1$ and  $n_2$ are smaller than $\frac \alpha {\alpha - 1}$, either both $G_1$ and $G_2$ are tree-like graphs so that $G$ is also a tree-like graph (and so that $G$ is not dominant if $n>\frac \alpha {\alpha - 1}$), or from the induction hypothesis, one of the $G_i$ satisfies $F_\alpha(G_i) < 1+\alpha + 2n_i $, so that  $$F_\alpha(G) < 2+2\alpha + 2(n_1 + n_2)  - (1+\alpha) = 1+\alpha + 2(n_1 + n_2),$$ which implies that $G$ is not dominant (tree-like graphs always have a larger $F_\alpha$).
 
\item If  $n_1<\frac \alpha {\alpha - 1}$ and  $n_2\ge \frac \alpha {\alpha - 1}$ (or conversely), we have  $$F_\alpha(G) \le 1+\alpha + 2n_1 + 1 + (\alpha + 1)n_2  - (1+\alpha) = 1+(\alpha + 1)(n_1 + n_2) + (1-\alpha) n _1,$$ so that $G$ is not dominant (star-like graphs always have a larger $F_\alpha$).

\item If both  $n_1$ and  $n_2$ are larger or equal to $\frac \alpha {\alpha - 1}$, we have  $$F_\alpha(G) \le 2+(\alpha+1)(n_1 + n_2)  - (1+\alpha)  < 1+(\alpha + 1)(n_1 + n_2) ,$$ so that $G$ is not dominant (star-like graphs always have a larger $F_\alpha$).

\end{itemize}

The only remaining case in this paragraph, is that for which the graph stays connected when exchanging the thin lines as in \eqref{fig:One-dotted-move}. In this case, we obtain a graph $G''$ with $\Fd(G) = \Fd(G'') - 1$ and either  $\Fs(G) = \Fs(G'')$ or $\Fs(G) = \Fs(G'') - 1$, depending on whether the solid face splits or not. In any case, we see that $F_\alpha(G)<F_\alpha(G'')$, so that $G$ is not dominant. This concludes the paragraph.

\

{\noindent \bf (c) \ On the existence of  solid faces with two trivalent solid nodes in a dominant graph. }
In this paragraph, we show that {\it if a graph contains a  solid faces with two trivalent solid nodes, then either $G$ is a star, or $G$ is not dominant (i.e.~we can find another connected graph with a larger $F_\alpha$)}. 

Since star-like graphs  are not dominant for $n<\frac \alpha{\alpha - 1}$, this implies that:
\begin{enumerate}[label=(\roman*), start=3]
\item \label{lemma21} {\it a dominant graph for $n<\frac \alpha{\alpha - 1}$ must satisfy $\Fs[(2)]=0$}.
\item \label{lemma22}{\it a graph with $n\ge \frac \alpha{\alpha - 1}$ for which $\Fs[(2)]>0$ is  a star-like graph or has a smaller $F_\alpha$ than star-like graphs.}
\end{enumerate}

Suppose that there exists a solid face in $G$ with two trivalent solid nodes. There are two possibilities locally, shown below (and all possible ways of crossing the three thin edges in the center for the graph on the left).  
\begin{figure}[h!]
\center
\includegraphics[scale=0.45]{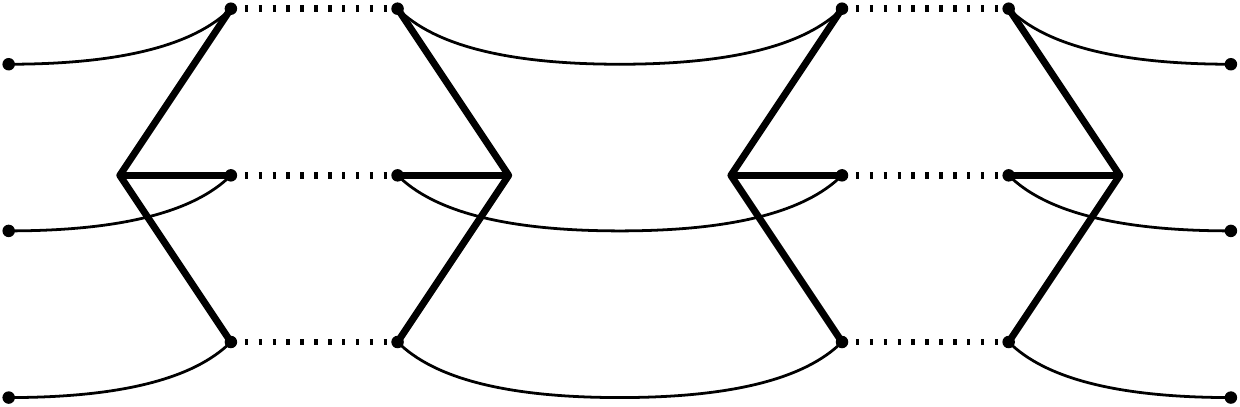}
\hspace{1.8cm}
\includegraphics[scale=0.45]{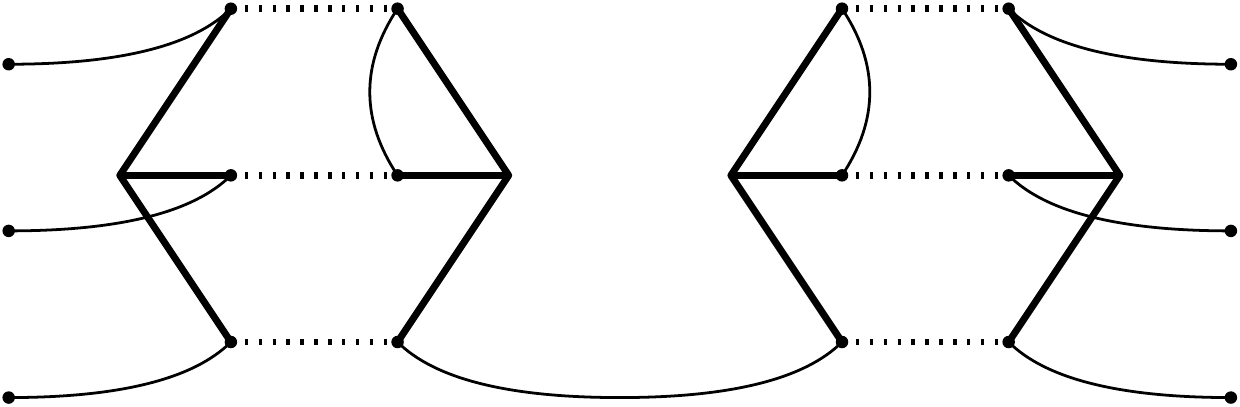}
\caption{Local possibilities for interactions around a solid face with two trivalent solid nodes}
\label{fig:Two-solid-cases}
\end{figure}

Let us first consider the case on the left of Fig.~\ref{fig:Two-solid-cases} (and possible crossings of the central thin edges), and perform the following move
\be
\label{fig:Two-solid-first-move}
\begin{array}{c}\includegraphics[scale=0.45]{Two-solid-1.pdf}\end{array}
\hspace{1cm}\rightarrow\hspace{1cm}
\begin{array}{c}\includegraphics[scale=0.45]{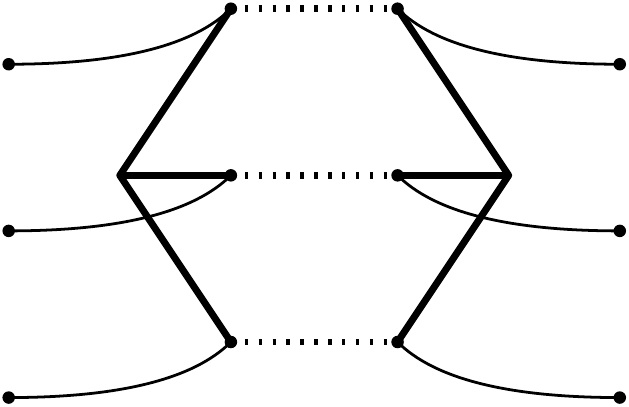}\end{array}
\ee 
in a way which respects the dotted faces. We obtain a graph $G'$ with the same number of dotted faces, and with one less solid face, so that $F_\alpha(G)=F_\alpha(G')+\alpha$. As usual, if $n-1<\alpha/(\alpha - 1)$, $F_\alpha(G')\le 1+\alpha+2(n-1)$, so that 
$$F_\alpha(G)\le 1+2\alpha+2n - 2 
= 1+ (\alpha + 1)n  + (n-2)(1-\alpha) < 1+ (\alpha + 1)n$$
as long as  $n>2$. 
On the other hand, if $n-1\ge \alpha/(\alpha - 1)$, $F_\alpha(G')\le 1+(\alpha+1)(n-1)$, so that 
$$F_\alpha(G)\le  1+(\alpha+1)n  - 1 
 < 1+ (\alpha + 1)n,$$
 so that the case on the left of Fig.~\ref{fig:Two-solid-cases} always leads to a non-dominant graph as long as $n>2$.

Let us now consider the case on the right of Fig.~\ref{fig:Two-solid-cases}. It is slightly more involved than the previous cases. First, let us specify that the results obtained for the move \eqref{fig:One-dotted-move} in the case where it disconnects the graph are slightly more general: consider a graph $G$ and two thin edges which belong to the same solid face, and such that  exchanging them disconnects the graph. Suppose in addition that one of these connected components is not a tree. Then the graph is not dominant, in the sense that we can always find graphs with larger $F_\alpha$. Indeed, the computations are precisely the same as what we have done before, with the difference that the case in which both $G_1$ and $G_2$ are tree-like graphs is excluded.

Let us consider the following move, which we can perform only if the two thin edges we exchange are indeed distinct. 
\be
\label{fig:Two-solid-last-move-0}
\begin{array}{c}\includegraphics[scale=0.45]{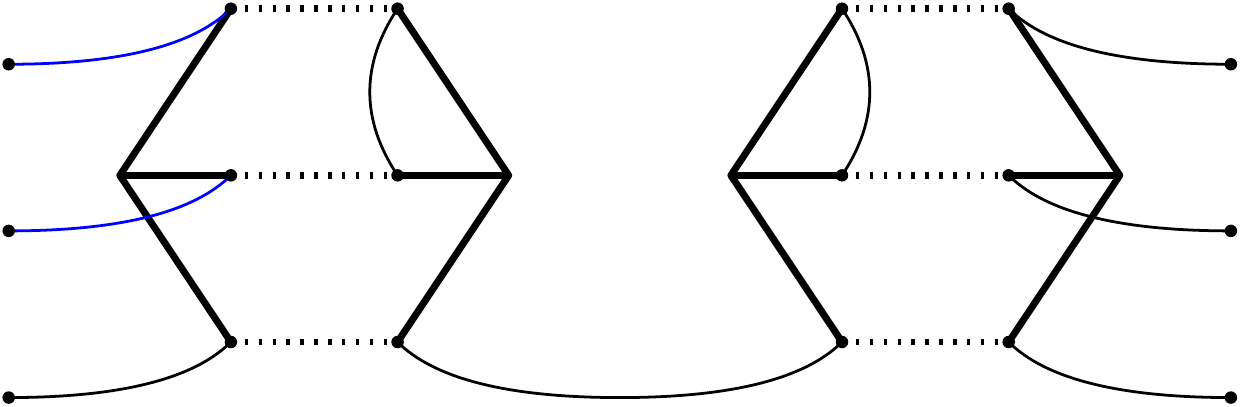}\end{array}
\hspace{1cm}\rightarrow\hspace{1cm}
\begin{array}{c}\includegraphics[scale=0.45]{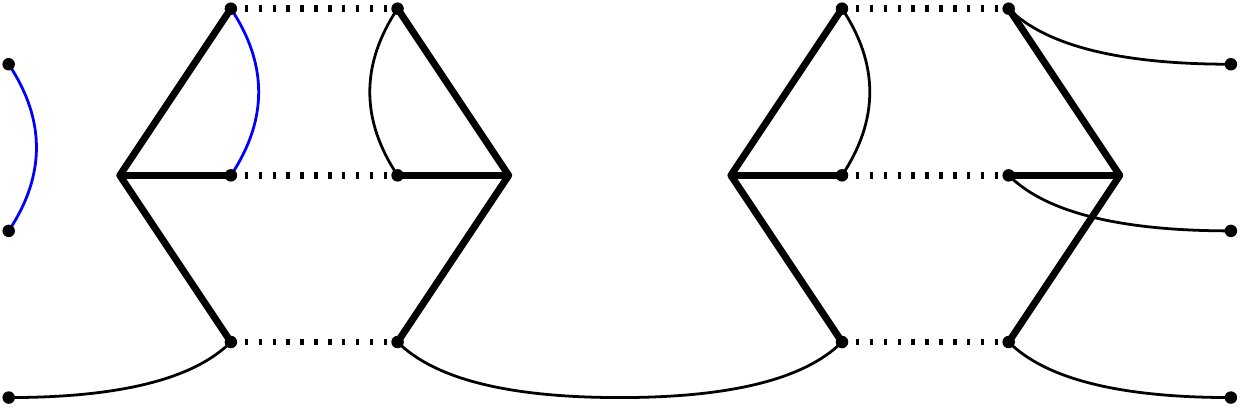}\end{array}.
\ee 
From what we just said, if it disconnects the graph, then $G$ is not dominant. If not, we obtain a connected graph $G''$, with one more dotted face, and either  one or zero additional solid face. Thus, $F_\alpha(G) < F_\alpha(G'')$. Therefore, a dominant graph $G$ with a solid face with precisely two trivalent solid nodes must be as follows,
\be
\label{fig:Two-solid-AB}
\begin{array}{c}\includegraphics[scale=0.45]{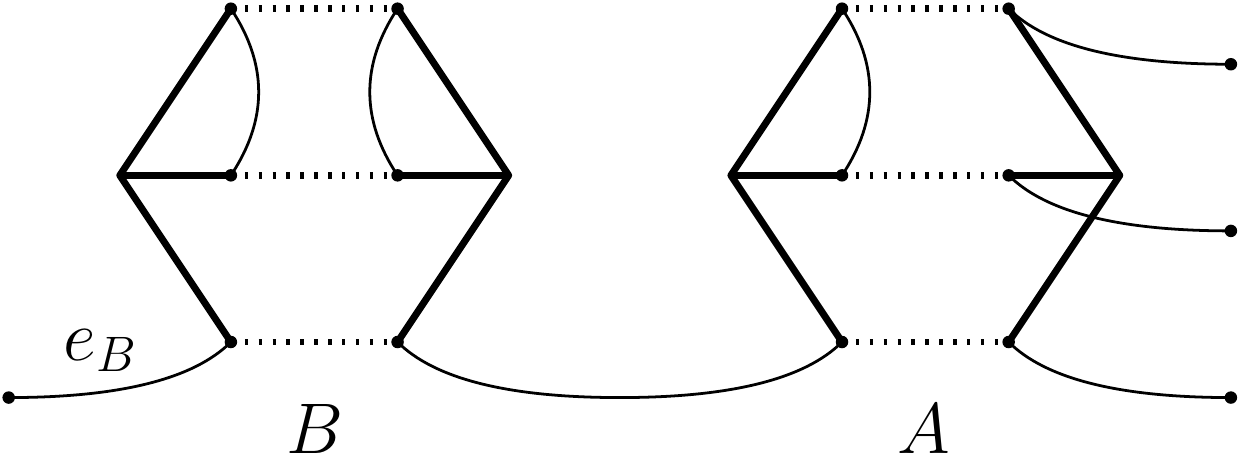}\end{array}.
\ee 

Let us now focus on the interaction $C$ attached to the other extremity of the thin edge $e_B$ on the left of  \eqref{fig:Two-solid-AB}, shown on the left of \eqref{fig:Two-solid-last-last-move} below (in the figure, we do not represent the interaction $A$ anymore). We perform the following move (if the two exchanged edges are distinct):
\be
\label{fig:Two-solid-last-last-move}
\begin{array}{c}\includegraphics[scale=0.45]{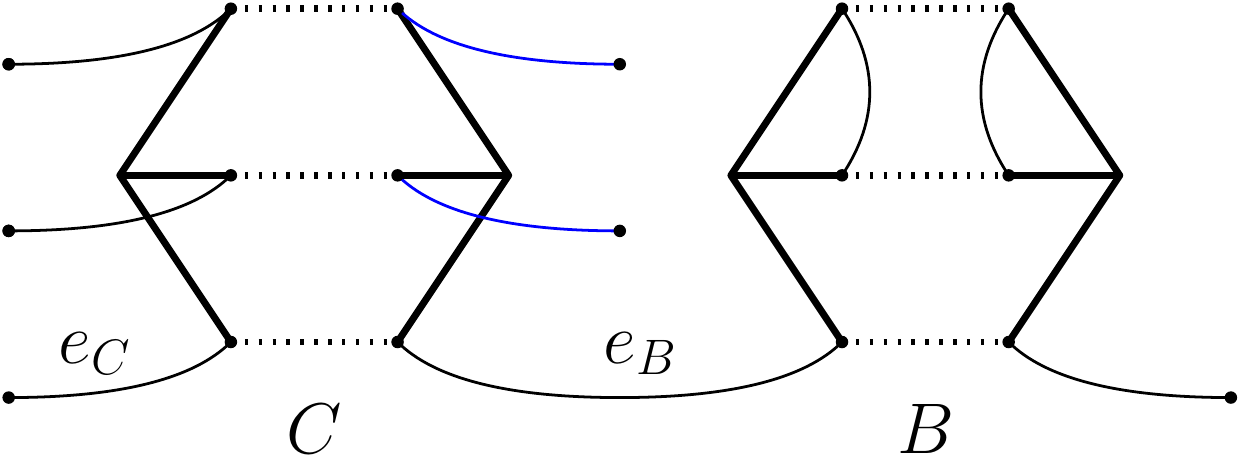}\end{array}
\hspace{1cm}\rightarrow\hspace{1cm}
\begin{array}{c}\includegraphics[scale=0.45]{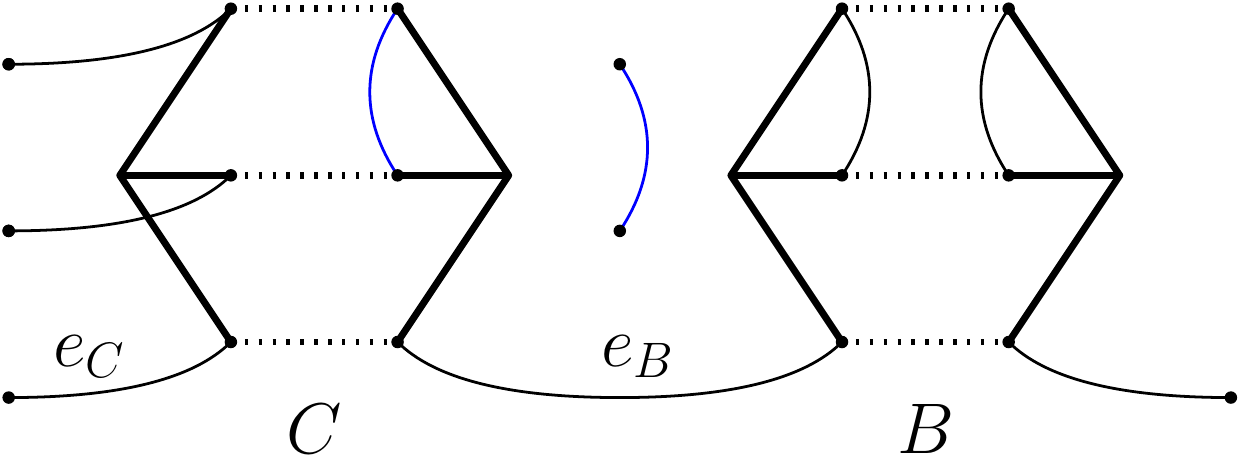}\end{array}.
\ee 
Applying the same argument again, we see that if this disconnects the graph $G$, then $G$ is not dominant, and if it does not disconnect the graph, it creates a solid face, while the number of dotted faces is modified by $-1$, $+1$, or 0. In any case, we obtain a graph $G'''$ with  $F_\alpha(G) \le F_\alpha(G''') + 1-\alpha < F_\alpha(G''')$. This means that for the graph to be dominant, the two edges we exchange in \eqref{fig:Two-solid-last-last-move} must in fact be the same edge, so that we are again in the situation on the left of \eqref{fig:Two-solid-last-move-0}, but for  the interactions $C$ and $B$ instead of $B$ and $A$.

 We then exchange the two thin edges on the upper left of $C$, concluding that if the graph is dominant, they must be the same edge, and we then focus on the interaction $D$ at the other extremity of $e_C$ and exchange the two thin edges on the upper right of the interaction $D$, concluding that if the graph is dominant, they must be the same edge, and so on. 
 
 We can apply repeatedly the moves \eqref{fig:Two-solid-last-move-0} and \eqref{fig:Two-solid-last-last-move}, to show that either the graph is not dominant, or it contains a larger and larger portion of star-like graph - a chain -  which we uncover from right to left at every step until eventually, all the $n$ interactions are included in the chain. This  forces the chain to be cyclic: when we uncover the $(n+1)$th interaction, the leftmost interaction must in fact be the rightmost interaction $A$. This proves that a  graph with a solid face with exactly two trivalent solid nodes is a star-like graph, or has a smaller $F_\alpha$ than some other graph. 

\

{\noindent \bf (d) \ Small  dominant graphs. } Since this is a proof by induction, we must study the cases for small $n$. 
We already know that \underline{for $n=1$}, the only dominant graph is the tree-like graph. 

\medskip

\underline{For $n=2$}, we have  $\Fs\in\{1,2\}$. If $\Fs=2$, we know that $\Fd\le 3$ with equality for the star-like graph (as is the case for larger $n$) or the other necklace graphs with the $R^2N^3$ behavior in \eqref{eq:dnr} (this is specific to $n=2$), in which case $F_\alpha= 3+2\alpha$.  If $\Fs=1$, we know that $\Fd\le 5$ with equality for the tree-like graph, in which case $F_\alpha= 5+\alpha$. If $\alpha<2$, the only dominant graph is the tree-like graph, while for $\alpha>2$, the only dominant graphs are the star-like graph and the other necklace graph. If $\alpha=2$, all of these graphs are dominant. 
However, this is  the only value of $n$ for which a graph which is not a tree-like graph or a star-like graph is dominant.

\medskip
\underline{For $n=3$}, there are three possibilities for the number of solid faces: $\Fs\in\{1,2,3\}$. Again, if $\Fs=3$ the graph is at best a star-like graph with $\Fd=4$ (now the only possible case), while if $\Fs=1$, the graph is at best a tree-like graph with $\Fd=7$. 

If $\Fs=2$, we know that $\Fd\le 6$ since the graph is not a tree-like graph. Let us suppose that $\Fd=6$ and $\Fs=2$ for a graph $G$. In that case, we use the lower bound on the number of dotted faces with a single dotted edge,  \eqref{eq:bound-small-dotted}, which implies that $\Fd[(1)]\ge 3$.
Suppose first that $\Fd[(1)]= 3$. Then  $\Fd[(l)]$ vanishes for $l>2$, so that from \eqref{eq:sum_lengths_faces}, $ \Fd[(1)] + 2\Fd[(2)]=9$, which implies that $\Fd[(2)]=3$. One can easily see that the graphs with $n=3$ and $\Fd[(1)]=\Fd[(2)]=3$ have $\Fs=1$. Therefore if $\Fs=2$ and $\Fd=6$, we must have $\Fd[(1)]\ge 4$.
Since there are three interactions, this implies that one of the interactions has two dotted faces with a single dotted edge each, i.e.~one of the interactions is as on the left of Fig.~\ref{fig:One-dotted-cases}. Applying the move \eqref{fig:One-dotted-last-move}, we have a graph $G'$ with two interactions and with two solid faces, so that at best there are three dotted faces. Thus at best, $\Fd(G) = \Fd(G')+2 = 5$ which contradicts the initial hypothesis that $\Fd(G)=6$. Thus at best, if $\Fs=2$, $\Fd=5$, so that $F_\alpha = 5 + 2\alpha$. 

We see that if $\alpha <3/2$, the tree-like graphs are dominant, while for $\alpha > 3/2$, only the star-like graph graph is dominant. For $\alpha=3/2$, both the star-graph and the tree-like graphs are dominant. This is the scenario we prove recursively, so that the case $n=3$ is enough to initiate the induction. 

\medskip

Using similar arguments, it is possible to prove that \underline{for $n=4$}, the maximum number of dotted faces at fixed number of solid faces are obtained for $R^4N^5$, $R^3N^6$, $R^2N^7$, and $RN^9$, but we do not detail the computations here.

\

{\noindent \bf (e) \ Proof of the results. }The proof is an induction on $n$. The results hold for $n=3$. Using the results above, under the induction hypothesis, we know from \ref{lemma21} that if $n<\frac \alpha{\alpha - 1}$, a dominant graph $G$ in $\bG_c(n)$ has $\Fs[(2)](G)=0$, so that from \eqref{eq:Bound-small-faces}, it must have $\Fd[(1)](G)>0$. We thus see from \ref{lemma12} that either $G$ is  a tree, or it has a smaller $F_\alpha$ than tree-like graphs. A dominant graph with $n<\frac \alpha{\alpha - 1}$ must therefore be a tree.

 Similarly, from \ref{lemma11}, if $n>\frac \alpha{\alpha - 1}$, a dominant graph $G$ in $\bG_c(n)$ has $\Fd[(1)](G)=0$, so that from \eqref{eq:Bound-small-faces}, it must have $\Fs[(2)](G)>0$. We therefore see from \ref{lemma22} that either $G$ is  a star, or it has a smaller $F_\alpha$ than star-like graphs . A dominant graph with $n>\frac \alpha{\alpha - 1}$ must therefore be a star.

If $n=\frac \alpha{\alpha - 1}$, both cases are possible, a graph $G$ in $\bG_c(n)$ cannot have both  $\Fd[(1)](G)=0$ and $\Fs[(2)](G)=0$, and we have shown that either it is a tree, either it is a star, either it is non-dominant.

\subsection{The large $R\sim N^\alpha$  regime with $\alpha\le 1$}
\label{sub:AlphaSmall1}
\subsubsection{Results}

In the previous section, we have shown that for $1<\alpha<2$, the dominant graphs were tree-like graphs for $n\le \frac\alpha{\alpha - 1}$, and star-like for $n\ge \frac\alpha{\alpha - 1}$. The dominant free-energy \eqref{eq:a-bet-12} consists of a polynomial of order $ \frac\alpha{\alpha - 1}$ corresponding to the tree-like graphs, and a series remainder corresponding to the star-like graphs. When $\alpha$ approaches 1, the polynomial part grows bigger and bigger, and we would  expect that  it would eventually  take over the full series when $\alpha\rightarrow 1^+$. In this section, we show that this is indeed the case, and that tree-like graphs are actually  dominant in the full domain  $0<\alpha\le 1$. 

\bigskip

To summarize, {\it in the large $R\sim N^\alpha$  regime with $\alpha\le 1$, dominant graphs are the tree-like graphs}, and the partition function, free energy, and two-point functions are given by those for finite $R$ and large $N$.

\bigskip

Note that although we do not know of any statistical physics interpretation to such a regime, from the point of view of the Feynman graph expansion of random coupling vector models \eqref{eq:random-coup-2} of large size $N$, this shows the robustness of the dominance of melonic graphs \cite{melons1, melons2, melons3} (Appendix \ref{app:melo}), since it remains valid when the number of replicas $R$ is large, but not larger than the size $N$ of the system.

\subsubsection{Proof}

We can adapt the proof of the previous section in that case, with a few modifications. It is again a recursion on $n$, initiated at $n=1$, for which we already know that the property holds.

\bigskip

{\noindent \bf Another bound on the number of small faces. }Again, we will need to develop  a similar lower bound as  \eqref{eq:Bound-small-faces} for the case where $\alpha \le 1$. We show that
 \be 
 \label{eq:Bound-small-faces-2}
G \in \bG_c(n)\text{ is dominant}\quad \Rightarrow \quad\Fd[(1)](G) + \alpha \Fs[(2)](G) \ge 2(\alpha + 1) + n(1 - \alpha ),
 \ee  
 where we recall that $\Fd[(l)]$ (resp.~$\Fs[(l)]$) is the number of dotted  (resp.~solid) faces incident to  $l$ interactions (counted with multiplicity). To prove this bound, we just use the lower bound \eqref{eq:small-bound-Falpha} on $F_\alpha$. Since we know that $F_\alpha^{\dom} \ge 1+\alpha + 2n$, a  dominant graph $G$ must satisfy
$$
\Fd[(1)](G)  + \alpha \Fs[(2)](G) \ge 2(1+\alpha+2n) -  n(3 +   \alpha),
$$
 which simplifies to \eqref{eq:Bound-small-faces-2}.
 
 \ 
 
 {\noindent \bf Dominant graphs have no solid faces with two trivalent solid nodes. }Let us consider a graph in $\bG_c(n)$ and suppose that $\Fs[(2)]>0$. Then $G$ must contain a subgraph as in Fig.~\ref{fig:Two-solid-cases}. For the case on the left of Fig.~\ref{fig:Two-solid-cases} (and possible crossings of the central thin edges), we perform the  move \eqref{fig:Two-solid-first-move}
in a way which respects the dotted faces. We obtain a graph $G'$ with the same number of dotted faces, and with one less solid face, so that $F_\alpha(G)=F_\alpha(G')+\alpha$. Using the induction hypothesis, $F_\alpha(G')\le 1+\alpha+2(n-1)$, so that 
$$F_\alpha(G)\le 1+\alpha+2n + ( \alpha - 2)
< 1+\alpha+2n,$$
so that $G$ is not dominant.

The case on the right of Fig.~\ref{fig:Two-solid-cases}  is slightly more involved, since the dotted faces are not conserved when we perform the following move:
\be
\label{fig:Two-solid-last-move}
\begin{array}{c}\includegraphics[scale=0.45]{Two-solid-2.pdf}\end{array}
\hspace{1cm}\rightarrow\hspace{1cm}
\begin{array}{c}\includegraphics[scale=0.45]{Two-solid-move.pdf}\end{array}.
\ee 
Upon performing this move, we suppress a dotted face if the two vertical thin edges belong to different dotted faces, and if they belong to the same dotted face, we either create  a dotted face or their number remains the same. In addition, a solid face is always suppressed. If $G''$ is the graph obtained after performing the move, we thus have $\Fd(G)=\Fd(G'')  + \eta$, where $\eta\in\{-1, 0, 1\}$, and $\Fs(G)=\Fs(G'') +1$. Importantly, if $G''$ is a tree, all three dotted edges on the right of \eqref{fig:Two-solid-last-move} lie in different dotted faces, so that if $G''$ is a tree-like graph we must have $\Fd(G)=\Fd(G'')  -1$. 
Using the recursion hypothesis,  $F_\alpha(G'') \le 1+\alpha + 2(n-1)$, with equality iff $G''$ is a tree,
in which case 
$F_\alpha(G) = 1+\alpha + 2(n-1) + \alpha - 1$.
Else if  $G''$ is not a tree,  $F_\alpha(G'') < 1+\alpha + 2(n-1) $ and $F_\alpha(G) < 1+\alpha + 2n-2 + \alpha +1 $. In both cases, 
$$F_\alpha(G)< 1+\alpha + 2n +\alpha -1    
<  1+\alpha + 2n,$$
 so that $G$ is not dominant. In conclusion, under the induction hypothesis, if $\Fs[(2)](G)>0$ and $\alpha\le1$, $G$ is not dominant. Using the bound \eqref{eq:Bound-small-faces-2} on small faces, this implies that a dominant graph $G$ must therefore satisfy $\Fd[(1)]>0$.

\

{\noindent \bf Dominant graphs are tree-like graphs. } Since a dominant graph must contain a dotted face with a single dotted edge, it must include a subgraph as in Fig.~\ref{fig:One-dotted-cases}. We review the various cases, everything works as before, with fewer cases. Again the cases in the middle of Fig.~\ref{fig:One-dotted-cases} are excluded in a dominant graph. 

If we have a  subgraph as on the left of Fig.~\ref{fig:One-dotted-cases}. Performing the move \eqref{fig:One-dotted-last-move}, we obtain a graph $G'$ with one less interaction and such that  $\Fd(G) = \Fd(G') + 2$ and $\Fs(G) = \Fs(G')$. Either $G'$ is a tree-like graph, in which case $G$ is a tree-like graph, or using the recursion hypothesis, $F_\alpha(G')<1+\alpha + 2(n-1)$, so that  $F_\alpha(G) < 1+\alpha + 2n$, in which case $G$ is not dominant.

If we have a subgraph as on the right of Fig.~\ref{fig:One-dotted-cases}, we perform the move \eqref{fig:One-dotted-move}. If the move disconnects the graph into two graphs $G_1$ and $G_2$, we have $F_\alpha(G) = F_\alpha(G_1) + F_\alpha(G_2)  - (1+\alpha)$.
Using the recursion hypothesis, either both $G_1$ and $G_2$ are tree-like graphs so that $G$ is also a tree-like graph, or one of the $G_i$ satisfies $F_\alpha(G_i) < 1+\alpha + 2n_i $, so that  $$F_\alpha(G) < 2+2\alpha + 2(n_1 + n_2)  - (1+\alpha) = 1+\alpha + 2(n_1 + n_2),$$ so that  that $G$ is not dominant.

If on the other hand the graph stays connected when exchanging the thin lines as in \eqref{fig:One-dotted-last-move}, as before we obtain a graph $G''$ with $\Fd(G) = \Fd(G'') - 1$
and either  $\Fs(G) = \Fs(G'')$ or $\Fs(G) = \Fs(G'') - 1$, depending on whether the solid face splits or not. In any case,  $F_\alpha(G)<F_\alpha(G'')$, so that $G$ is not dominant. This concludes the proof.

\section{Describing the model in a convergent series}
\label{sec:convergent}
The expansion in $\lambda$  of the partition function $Z_{N,R}(\lambda,k)$ defined in \eq{eq:integral} does not give a convergent
series in general,
because there exists an essential singularity at $\lambda=0$. 
This is obvious from the form of \eq{eq:integral},
because the integral diverges 
for $\lambda<0$. 
In fact, the situation can explicitly be checked in the exactly solvable case $R=1$, 
as we will see in Section~\ref{sec:Req1}. 
The reason why we obtained the convergent results 
in Section~\ref{sec:graph} comes from the fact that we 
summed up  the dominant graphs only\footnote{This is usually the case for vector, matrix and tensor models.}. 
In this section, to have more control over the situation,
we will divide the integral over $\phi_a^i$ into its angular and  radial parts. 
We will see that the angular part admits a convergent series expansion in $\lambda$,
whose coefficients are expressed in terms of the coefficients $z_n$ of the 
Feynman diagrammatic expansions of Section~\ref{sec:graph}.
On the other hand, the radial part will be treated in a different manner as an explicit integration.
We will finally apply our results to 
discuss the integrability of the wave function of a toy model 
\cite{Obster:2017pdq}  closely related to the tensor model 
in the Hamilton formalism introduced in \cite{Sasakura:2011sq,Sasakura:2012fb}.

\subsection{Dividing the integration into the angular and radial parts}
\label{sec:probability}
Let us break $\phi_{a}^i$ into the radial part $r^2:=\Tr\phi \phi^t$
and
the angular part $\tilde \phi_a^i:=\phi_a^i/r$, which represents coordinates on a unit sphere, $S^{NR-1}$.
Then, one can rewrite $Z_{N,R}$ in \eq{eq:integral} as
\[
Z_{N,R}(\lambda,k) =\hbox{vol}\left(S^{NR-1}\right)
\int_0^\infty dr\, r^{NR-1}   f_{N,R}(\lambda r^6) e^{-k r^2},
\label{eq:zwithf}
\]
where
\[
f_{N,R}(t):=\frac{1}{\hbox{vol}\left(S^{NR-1}\right)}\int_{S^{NR-1}}d\tilde \phi\ 
\exp\left(-t\, 
U(\tilde \phi) \right)
\label{eq:defoff}
\]
with $U$ defined in \eq{eq:Interaction} and $\hbox{vol}\left(S^{NR-1}\right)$ denoting the volume of the unit sphere, $\int_{S^{NR-1}}d\tilde \phi$.
For finite  $N,R$ and for complex $t$, 
$f_{N,R}(t)$ is an entire function\footnote{i.e.~it is holomorphic at every finite point of $\mathbb{C}$.}
because of the form of \eq{eq:defoff}, which is an integration of an exponential function of $t$ over a compact space.

As an entire function, it is differentiable over $\bR$, and since 
\be 
\label{eq:non-negative-interaction}
U(\tilde \phi)=
\sum_{i,j=1}^R\left(\sum_{a=1}^N \tilde \phi^i_a \tilde \phi^j_a\right)^3
=\sum_{a,b,c=1}^N
\left(\sum_{i=1}^R  \tilde \phi_a^i
\tilde \phi_b^i\tilde \phi_c^i\right)\left(\sum_{j=1}^R \tilde \phi_a^j\tilde \phi_b^j\tilde \phi_c^j\right)\ge 0,
\ee
we see that $f_{N,R}(t)$ is a monotonically decreasing positive function for  finite $N,R$ and real $t$ with $f_{N,R}(0)=1$.

Furthermore, as an entire function, the series expansion of $f_{N,R}$ in $t$ 
is convergent, and the radius of convergence is infinite.
In Section~\ref{sec:pertf}, we will express the coefficients of the series expansion of $f_{N,R}$ in terms of the coefficients $z_n$, which are computed using Feynman graphs as detailed in the Section~\ref{sub:Feyn}.

The angular part of the integration is contained 
in the expression \eq{eq:defoff} of $f_{N,R}$.
This integration is over a compact space and free from the 
variables $\lambda,k$, but it is still highly non-trivial.
It is closely related to what  appears
in the $p$-spin spherical
model \cite{pspin,pedestrians} for 
the spin glass,
as the integration variables are constrained to be on a unit sphere.
Therefore we would be able to apply 
to our model the various techniques 
which have been
developed for the understanding of this spin glass model. 
Although we  rather use diagrammatic expansions 
in the present paper, such applications
would be of potential interest.

To interpret $f_{N,R}$, we express it as the moment-generating function (or Laplace transform)
\be
f_{N,R}(t)=\int_0^1 d\sigma\, \rho_{N,R}(\sigma)  \exp(-t\, \sigma),
\label{eq:relfrho}
\ee
of the following probability density,
\[
\rho_{N,R}(\sigma):=\frac{1}{\hbox{vol}\left(S^{NR-1}\right)}\int_{S^{NR-1}}d\tilde \phi\ \delta \left(
\sigma- U(\tilde \phi)
\right).
\label{eq:defofrho}
\]
This quantity obviously satisfies  $\rho_{N,R}(\sigma)\geq 0$ and $\int d\sigma\, \rho_{N,R}(\sigma)=1$, which justifies that it can indeed be regarded as a probability density over $\sigma$.
Furthermore, we see from \eqref{eq:non-negative-interaction} and 
\[
U(\tilde \phi)=
\sum_{i,j=1}^R\left(\sum_{a=1}^N\tilde \phi^i_a \tilde \phi_a^j\right)^3 
\leq
\sum_{i,j=1}^R \left(\sum_{a,b=1}^N \tilde \phi^i_a\tilde \phi_a^i  \tilde \phi^j_b\tilde \phi^j_b
\right)^\frac{3}{2}\leq \sum_{i,j=1}^R
\sum_{a,b=1}^N\tilde \phi^i_a\tilde \phi^i_a  \tilde \phi^j_b \tilde \phi^j_b=1,
\label{eq:ineqforint}
\]
that the support of $\rho_{N,R}(\sigma)$ is included in $0\leq\sigma\leq1$. In \eqref{eq:ineqforint}, we have used the Cauchy-Schwartz inequality, $\sum_{a=1}^N \tilde \phi^i_a \tilde \phi^j_a\leq \left(\sum_{a,b=1}^N\tilde \phi^i_a \tilde \phi^i_a
\,\tilde \phi^j_b\tilde \phi^j_b\right)^\frac{1}{2}$, and $0 \leq  \sum_{a=1}^N \tilde\phi^i_a \tilde \phi^i_a\leq 1$.

In terms of $\rho_{N,R}$, the partition function
is expressed as
\[
Z_{N,R}(\lambda,k)=
\hbox{vol}\left(S^{NR-1}\right) \int_0^1 d\sigma \, \rho_{N,R}(\sigma) \int_0^\infty dr \, r^{NR-1} e^{- \lambda \sigma r^6-k r^2}.
\label{eq:zwithrho}
\]
In this expression,
 the angular part $\rho_{N,R}$ and 
 the radial part can be treated independently, and they are combined by
 the last integration over $\sigma$.
As in the case of $f_{N,R}$, the angular integration for $\rho_{N,R}$ is highly non-trivial.
 On the other hand, the integration over $r$
 is rather straightforward, 
and one can obtain an explicit expression in terms of the generalized hypergeometric function 
$_1F_2$, as shown in Appendix~\ref{app:explicit}.
An important property is that  it has an essential singularity at $\lambda=0$ (as the partition function $Z_{N,R}(\lambda,k)$), 
which is consistent with the fact that  the series expansion of $Z_{N,R}(\lambda,k)$ in $\lambda$  is not convergent.

The probability density $\rho_{N,R}$ also has an interesting meaning in 
the context of tensor-rank decomposition
(or CP-decomposition) 
in computer science
\cite{SAPM:SAPM192761164,comon:hal-00923279,Landsberg2012},
 which is an important technique for analyzing tensors 
representing data.
This technique decomposes a tensor,
 say a symmetric tensor $Q_{abc}\ (a,b,c=1,2,\ldots,N)$, 
 into a sum of rank-one tensors as
\be
Q_{abc}=\sum_{i=1}^R \phi_a^i \phi_b^i \phi_c^i ,
\label{eq:tensordec}
\ee
where $R$ is called the rank of $Q_{abc}$ (more precisely, for a given tensor, its rank is the smallest $R$ which realizes 
such a decomposition).
It is not well understood
how such rank-$R$ tensors 
exist in the space of all tensors, especially for real tensors.
Since $\sum_{a,b,c=1}^N
Q_{abc}Q_{abc}=U(\phi)$,
the probability density $\rho_{N,R}$ in \eq{eq:defofrho}
gives 
a part of such knowledge, namely, 
the size distributions of tensors with
a certain rank under the normalization
$\Tr\phi \phi^t=1$.

\subsection{The series expansion 
of the angular part}
\label{sec:pertf}
In this subsection, we will compute the series 
expansion of $f_{N,R}(t)$ in
$t$, which is guaranteed to have an infinite convergent radius,
as discussed in Section~\ref{sec:probability}. This could be applied to other models as well.

By performing the Taylor expansion of $f_{N,R}(t)$ in \eq{eq:defoff} in $t$, one obtains
\[
f_{N,R}(t)&= \sum_{n=0}^\infty (-t)^n C_{N,R}(n),
\label{eq:expandf}
\]
where
\[
C_{N,R}(n)&=\frac{(-1)^n}{n!} \left. \frac{d^n}{dt^n}f_{N,R}(t) \right|_{t=0} \CR
&=\frac{1}{n!} \frac{1}{\hbox{vol}\left(S^{NR-1}\right)}
\int_{S^{NR-1}}d\tilde \phi\ \left(
U(\tilde \phi)
\right)^n. 
\]
Here, 
changing
the order of the derivative and the integration
is allowed for this
well-behaved integration.
For any arbitrary positive  constant  $\beta$, we have 
\[
C_{N,R}(n)&=\frac{1}{n!}\frac{ \int_{\mathbb{R}^{NR}} d\phi 
\left(\frac{U(\phi)}{\left(\Tr\phi \phi^t\right)^3 }\right)^n
\exp \left( -\beta\Tr\phi \phi^t\right)}{\int_{\mathbb{R}^{NR}} d\phi 
\exp \left( -\beta \Tr\phi \phi^t\right)}.
\label{eq:defofC}
\]
Indeed, introducing a radial direction by $\phi_a^i=\tilde \phi_a^i r$, we see that the integrations over $r$ cancel between the numerator 
and the denominator, and
$\beta$ is indeed a dummy variable, which does not appear in the final expression of $C_{N,R}$.
In particular,
$\beta$ has nothing to do with the parameter
$k$ in \eqref{eq:zwithf}.

The numerator in the last line of \eq{eq:defofC} has the following obvious properties: on one hand, by performing the rescaling $\phi\rightarrow \phi/\sqrt{\beta}$ we see that 
\[
 \int_{\mathbb{R}^{NR}} d\phi &
\left(\frac{U(\phi)}{\left(\Tr\phi \phi^t\right)^3 }\right)^n
e^{ -\beta\Tr\phi \phi^t}  = \beta^{-\frac{NR}2} A, 
\label{eq:A}
\]
where $A$ does not depend on $\beta$, while on the other hand, 
\[
(-1)^{3n}\frac{d^{3n}}{d\beta^{3n}} \int_{\mathbb{R}^{NR}} d\phi &
\left(\frac{U(\phi)}{\left(\Tr\phi \phi^t\right)^3 }\right)^n
e^{ -\beta\Tr\phi \phi^t} 
=\int_{\mathbb{R}^{NR}} d\phi 
\left(U(\phi)\right)^n
e^{ -\beta\Tr\phi \phi^t},
\label{eq:derbeta}
\]
which is equal to $(\frac {\pi}{\beta})^{\frac{NR}2} n ! z_n(N,R,\beta)$, where $z_n({N,R,\beta})$ are the expansion coefficients of 
the partition function defined in \eq{eq:Part-Func-Exp-1}.
Differentiating \eqref{eq:A}, we determine $A$ 
and obtain the following relation,
\[
C_{N,R}(n)=\frac{\Gamma\left(\frac{NR}{2}\right)\beta^{3n}}{\Gamma\left(\frac{NR}{2}+3n\right)} \,
z_n({N,R,\beta}).
\label{eq:relofCandd}
\]
Since
$z_n({N,R,\beta})=z'_n(N,R)/{(8\beta^3)^n}$ (see below \eq{eq:Part-Func-Exp-1}), the 
dummy parameter $\beta$  cancels
out from the expression.

The relation \eq{eq:relofCandd}
provides a method for determining the series
expansion of the angular part from the standard series expansion with the Feynman graphs. As for the $n$-dependence, \eq{eq:relofCandd} shows that $C_{N,R}(n)$ decays much faster than $z_n(N,R,\beta)$ in $n$. 
Therefore, in general, 
$f_{N,R}$ has a much faster convergent series than that of the partition function.
This is consistent with the argument made in Section~\ref{sec:probability} 
that $f_{N,R}$ is an entire function, while the partition function is not.

\subsection{The $R=1$ example}
\label{sec:Req1}
The behaviors
of $f_{N,R}$ and the partition function $Z_{N,R}$ 
mentioned in Section~\ref{sec:pertf} 
can explicitly be checked in the trivial solvable case with $R=1$.\footnote{This case 
corresponds to a one-vector model
\cite{Nishigaki:1990sk,DiVecchia:1991vu}
with a sixth order interaction term.}
In this case, 
$U(\tilde \phi)=(\Tr\tilde\phi \tilde\phi^t)^3=1$ identically
 from the normalization of 
 $\tilde \phi$,
and hence from \eq{eq:defofC}, we obtain
\[
C_{N,1}(n)=\frac{1}{n!}.
\label{eq:CforReq1}
\]
Then the series  \eq{eq:expandf} can be summed up to
\[
f_{N,1}(t)=e^{-t}.
\]
As mentioned in Section~\ref{sec:probability}, this 
is in fact an entire function of $t$.
By putting it into \eq{eq:zwithf}, one obtains 
\[
Z_{N,1}(\lambda,k)=\hbox{vol}\left(S^{N-1}\right)
\int_0^\infty dr\, r^{N-1}   e^{-\lambda r^6-k r^2}.
\label{eq:zreq1}
\]
This is of course equivalent to what one will obtain by directly parametrizing 
$\phi_a^{i=1}$  with the radial and 
angular coordinates in the original expression \eq{eq:integral}, and integrating out the trivial angular part. 
The remaining integration over $r$ can be expressed using 
the hypergeometric function ${}_1F_2$ as derived in Appendix~\ref{app:explicit}.
The result has an essential singularity 
at $\lambda=0$, as expected.

This last statement 
can also be checked from the 
explicit form of $z_n$.
We obtain
\[
z_n(N,1,k)=\frac{\Gamma\left(\frac{N}{2}+3n\right)}{n!\  \Gamma\left(\frac{N}{2}\right)} k^{-3n}. 
\]
This can be obtained from the relation \eq{eq:relofCandd} with $\beta=k$ and \eq{eq:CforReq1}, 
or even directly computing the Gaussian integration in \eq{eq:Part-Func-Exp-1}.
This is a divergent series and 
its interpretation is not 
straightforward, as widely discussed in the literature on random vector models.

As shown in this trivial 
$R=1$ example,
it seems useful to 
divide the integration 
into the angular and radial
directions for explicit evaluations of 
the partition function $Z_{N,R}(\lambda,k)$,
rather than directly 
treating a highly divergent
series in $\lambda$
with $z_{n}(N,R,k)$.

\subsection{Application to a tensor model}
\label{sec:limit}
In this subsection, 
we will apply the results of the previous subsections
to study the integrability of the wavefunction of the model introduced in \cite{Obster:2017pdq}. It is a toy model  closely related to a tensor model 
in the Hamilton formalism, called the canonical tensor model \cite{Sasakura:2011sq,Sasakura:2012fb}, which is studied in a quantum gravity context.

Let us  
consider the following wave function
depending on a symmetric tensor $P_{abc}$, where $\ (a,b,c=1,2,\ldots,N)$:
\[
\psi(P):=\int_{\mathbb{R}^N} d\varphi \exp\left(I
\sum_{a,b,c=1}^N P_{abc}\varphi_a\varphi_b\varphi_c
+\left(I-\epsilon\right)
\sum_{a=1}^N\varphi_a \varphi_a\right),
\label{eq:psip}
\]
 where $I$ denotes the imaginary unit $I^2=-1$, and $d\varphi:=\prod_{a=1}^N d\varphi_a$.
 For general real $P_{abc}$, the integral \eq{eq:psip} is oscillatory
 and is regularized by 
 a small positive regularization parameter $\epsilon$ 
of the so-called Feynmann prescription,
in which $\epsilon\rightarrow +0$ is supposed to be lastly taken.

In \cite{Obster:2017pdq},
it was argued and explicitly shown for 
some simple cases that
the wave function \eq{eq:psip} 
has coherent peaks
for some specific loci of $P_{abc}$ where
$P_{abc}$ is invariant under Lie-group transformations
 (namely, $P_{abc}=h_a^{a'}h_b^{b'}h_c^{c'}P_{a'b'c'}$
 for  $^\forall h\in H$ with a Lie-group representation $H$). 
 In fact, a tensor model \cite{Ambjorn:1990ge,Sasakura:1990fs,Godfrey:1990dt} 
 in the Hamilton formalism \cite{Sasakura:2011sq,Sasakura:2012fb}
 has a similar wave function $\tilde \psi(P)^R$ with
a power $R$ and $\tilde \psi(P)$ very similar to $\psi(P)$ \cite{Narain:2014cya},
and it was shown in \cite{Obster:2017dhx} that the wave function of this tensor model has similar coherent peaks. 
To consistently interpret this 
phenomenon as the preference for
Lie-group symmetric configurations in the tensor model, we first have to show that 
we can apply the quantum mechanical 
probabilistic interpretation to the wave
function, namely,
the wave function must be absolute 
square integrable.
This is a difficult question even for the toy wave function \eq{eq:psip}, since it
has complicated dependence on $P_{abc}$ mainly
due to its oscillatory character.

As a first step towards answering this question,
in this paper we will study the behavior of the following quantity in $\kappa$:
\[
g(N,R,\kappa):=\int_{\mathbb{R}^{\#P}} dP\ \exp\left( -\kappa \sum_{a,b,c=1}^N P_{abc}P_{abc} \right) \psi(P)^R,
\label{eq:defofg}
\]
where $\#P:=N(N+1)(N+2)/6$ is  the number of 
independent components of the symmetric tensor $P_{abc}$, and 
 $dP:=\prod_{\genfrac{}{}{0pt}{3}{a,b,c=1}{a\leq b\leq c}}^N \sqrt{d_{abc}}\,dP_{abc}$
 with a degeneracy factor, $d_{aaa}=1,\ d_{aab}=3$, and  $d_{abc}=6$
 for $a<b<c$.
In the $\kappa\rightarrow +0$ limit,
this quantity coincides with the integration of the wave function $\psi(P)^R$ over 
the whole space of $P_{abc}$. If this is finite
in the limit, the wave function is integrable.
We may regard this as a toy case study 
towards proving the square-integrability
of the wave function \cite{Narain:2014cya, Obster:2017dhx}
in the canonical tensor model of \cite{Sasakura:2011sq,Sasakura:2012fb}.
By putting \eq{eq:psip} into \eq{eq:defofg}
and integrating over $P_{abc}$, we obtain
\[
g(N,R,\kappa)&=\int_{\mathbb{R}^{\#P} }dP\int_{\mathbb{R}^{NR}} d\phi \exp \left( -\kappa \sum_{a,b,c=1}^N P_{abc}P_{abc}+I \sum_{i=1}^R \sum_{a,b,c=1}^N P_{abc}\phi_a^i\phi_b^i\phi_c^i \right. \CR
&\hspace{5cm}\left.+(I-\epsilon) \sum_{i=1}^R \sum_{a=1}^N\phi^i_a\phi^i_a\right)
\CR
&=\left(\frac{\pi}{\kappa}\right)^{\frac{\#P}{2}} Z_{N,R}\left( \frac{1}{4 \kappa},-I+\epsilon\right),
\label{eq:gtoz}
\]
where $Z_{N,R}$ is the partition function of our matrix model
\eq{eq:integral}.
 
By using \eq{eq:gtoz} and \eq{eq:zwithrho}, $g(N,R,\kappa)$ can also be expressed as
\[
g(N,R,\kappa)=\hbox{vol}\left(S^{NR-1}\right) \left(\frac{\pi}{\kappa}\right)^{\frac{\#P}{2}} \int_0^1 d\sigma \, \rho_{N,R}(\sigma) \int_0^\infty dr \, r^{NR-1} e^{- \sigma r^6/(4\kappa)+(I-\epsilon) r^2}.
\]
From this expression, one can see that the most delicate region of the integration over $\sigma$ is located 
near the origin, since the integration over $r$ may need careful treatment in the large $r$ region
for $\sigma\sim +0$. 
In addition, the $\sigma\sim +0$ region becomes more important, as $\kappa$ is taken smaller
for our interest in the $\kappa\rightarrow +0$ limit.
Therefore, 
it is essentially important to determine the behavior of $\rho_{N,R}(\sigma)$ near the origin. 
In turn, from the relation \eq{eq:relfrho}, 
this is equivalent to determining the $t\rightarrow +\infty$ behavior of $f_{N,R}(t)$. This can also be seen directly from
\[
g(N,R,\kappa) =\hbox{vol}\left(S^{NR-1}\right)  \left(\frac{\pi}{\kappa}\right)^{\frac{\#P}{2}}
\int_0^\infty dr\, r^{NR-1}   f_{N,R}\left(\frac{r^6}{4 \kappa }\right) e^{(I-\epsilon) r^2},
\label{eq:gwithf}
\]
which can be obtained by expressing $g(N,R,\kappa)$ with $f_{N,R}$ 
using \eq{eq:zwithf}.

While the toy model of \cite{Obster:2017pdq} allows any value of $R$,
the tensor model  \cite{Sasakura:2011sq,Sasakura:2012fb}
uniquely requires $R=(N+2)(N+3)/2$ \cite{Obster:2017dhx,Sasakura:2013wza}
for the hermiticity of the Hamiltonian.\footnote{The wave function of the tensor model is given by $\psi(P)^{\lambda_H/2}$ with $\lambda_H=(N+2)(N+3)/2$
\cite{Obster:2017dhx,Sasakura:2013wza}. Our interest in this paper is its square integrability, which is a toy case study for the absolute square integrability.
Therefore, $R= \lambda_H$. }
Because of the very similar form of the wave functions of the models, our interest is therefore especially in the regime $R \sim N^2$.
The dominant graphs at large $N$ at each order in $\lambda$ for $R\sim N^\alpha$ have systematically been analyzed in Section~\ref{sub:AlphaLarg1}, and it has been found that there exists a transition region 
$1\leq \alpha \leq 2$, where the dominant graphs gradually change.
This would imply that the dynamics of the model 
are largely different between the two regions
$R\gtrsim N^2$ and $R \lesssim N$.
Motivated by this fact, 
we compute $f_{N,R}$ through
the relation \eq{eq:relofCandd}
first by using the result in Section~\ref{sub:LargeRfiniteN}, which incorporates all the necklace graphs and is a valid
approximation for large $R$ and finite $N$. 
We will also comment on 
how our result will change if we take  \eq{eq:allarger2} and \eq{eq:aleq2},
 which 
 come from  the dominant graphs
in large $N$ for $\alpha>2$ 
and $\alpha=2$,
respectively.

In the leading order of large $NR$, which includes the regime of large $R$ and finite $N$, and also all the other regimes with $R\sim N^\alpha$ discussed in Section~\ref{sec:graph}, the relation \eq{eq:relofCandd} is given by
\be
C_{N,R}(n)_{leading}=\left( \frac{NR}{2}\right)^{-3n} \beta^{3n} \, z_n(N,R,\beta)_{leading},
\label{eq:cnrlead}
\ee
where we have formally employed the following expansion in $1/NR$,
\[
\frac{\Gamma\left(\frac{NR}{2}\right)}{\Gamma\left(\frac{NR}{2}+3n\right)} =
\prod_{i=0}^{3n-1} \left( \frac{NR}{2}+i\right)^{-1}=\left( \frac{NR}{2}\right)^{-3n}\left(1+O\left((NR)^{-1}\right) \right),
\label{eq:formal}
\]
and $z_n(N,R,\beta)_{leading}$ denotes the leading order of the 
coefficient $z_n(N,R,\beta)$ in any of the regimes discussed in Section~\ref{sec:graph}. 
Note that here we have 
assumed the existence of a $1/R$
expansion or
other expansions with $R\sim N^\alpha$ 
for $C_{N,R}(n)$, or $f_{N,R}$,
to employ 
the formal expansion
in $1/NR$
 irrespective of the value of $n$ in \eq{eq:formal}.

In the regime of large $R$, by using the result from Section~\ref{sub:LargeRfiniteN} one obtains
\[
f_{N,R}(t)_{leading}&=\sum_{n=0}^\infty (-t)^n C_{N,R}(n)_{leading} \CR
&=\sum_{n=0}^\infty \left(-\frac{8\beta^3 t}{N^3R^3}\right)^n z_n(N,R,\beta)_{leading} \CR
&=\left( 1+ \frac{6(N+4)t}{N^3 R^2}\right)^{-\frac{N}{2}}
\left(
1+ \frac{12t}{N^3 R^2}
\right)^{-\frac{N(N+4)(N-1)}{12}},
\label{eq:fleading}
\]
where we have put \eq{eq:cnrlead} into \eq{eq:expandf} and have used \eq{eq:resultsum}.
As can be seen in \eq{eq:fleading}, the $f_{N,R}(t)_{leading}$ has an interesting scaling property 
at large $t$, which will become important in the analysis below.

 Let us 
 check \eq{eq:fleading}
from the point of view of 
the expected properties of $f_{N,R}(t)$.
 As explained in Section~\ref{sec:probability}, 
 $f_{N,R}(t)$ should be a monotonically decreasing function for real $t$ with $f_{N,R}(0)=1$.
 This is satisfied by  \eq{eq:fleading} in the region $t\geq 0$, which is the integration region for 
 the computation of $g(N,R,\kappa)$ 
 as in \eq{eq:gwithf}.
On the other hand, $f_{N,R}(t)$ should be an entire function of $t$ as explained in Section~\ref{sec:probability}. This is not satisfied by \eq{eq:fleading}, as there exist singular points 
at $t\sim -N^2R^2,\ -N^3 R^2$.
Considering the fact that 
we are discussing
the large-$R$ regime, 
the singular points can be regarded as 
being far away from the 
integration region $t\geq 0$.
However,
for $f_{N,R}(t)$ to be an entire function,
these 
singularities in \eq{eq:fleading} must be canceled by
the sub-leading corrections of $f_{N,R}(t)$ to \eq{eq:fleading}.
This means that 
the sub-leading corrections
may 
also be a series whose
radius of convergence is of the order of $N^2 R^2$.
Therefore,
$f_{N,R}(t)$ may get 
some important 
corrections at $t\gtrsim N^2 R^2$
from such sub-leading 
contributions (see also Section~\ref{sec:discussionlimit}).
Though this should be taken as a caution in using \eq{eq:fleading},
we will use it beyond this limit in the computation below
as the leading order expression of the entire function $f_{N,R}(t)$.

By putting \eq{eq:fleading} into \eq{eq:gwithf}, we obtain
\[
\begin{split}
&g(N,R,\kappa)_{leading}\\ &\ \ =\hbox{vol}\left(S^{NR-1}\right) \left(\frac{\pi}{\kappa}\right)^{\frac{\#P}{2}} \\
&\ \ \ \ \ 
 \times \int_0^\infty dr\, r^{NR-1}  \left( 1+ \frac{3(N+4)r^6}{2\kappa N^3 R^2}\right)^{-\frac{N}{2}}
\left(
1+ \frac{3 r^6}{\kappa N^3 R^2}
\right)^{-\frac{N(N+4)(N-1)}{12}}e^{(I-\epsilon) r^2}.
\end{split}
\]
From now on, let us concentrate only on the behavior in $\kappa$. 
By the change of variable, $r\rightarrow \kappa^{1/6}r$, we obtain
\[
&g(N,R,\kappa)_{leading} \CR
&\ \ \propto \kappa^{-\frac{\#P}{2}+\frac{NR}{6}}
\int_0^\infty dr\, r^{NR-1}  \left( 1+ \frac{3(N+4)r^6}{2N^3 R^2}\right)^{-\frac{N}{2}}
\left(
1+ \frac{3 r^6}{ N^3 R^2}
\right)^{-\frac{N(N+4)(N-1)}{12}}e^{(I \kappa^{1/3}-\epsilon)r^2}.
\label{eq:gwithfresult}
\]
We now divide further discussions into the following three cases.

\noindent
(i)  $R < (N+1)(N+2)/2$\\
\noindent
In this case, the $\kappa\rightarrow +0$ 
behavior of the integration in \eq{eq:gwithfresult} 
converges to a finite non-zero value, 
because the modulus of the integrand damps fast enough in $r$.
Therefore, the behavior of $g(N,R,\kappa)_{leading}$ is
determined by the factor in front.  
By putting $\#P=N(N+1)(N+2)/6$ (see below \eq{eq:defofg}),
we obtain
\[
g(N,R,\kappa)_{leading}\sim \kappa^{\frac{N}{6}\left(R-\frac{(N+1)(N+2)}{2}\right)}.
\label{eq:gcasei}
\]
Therefore it has diverging behavior in the limit $\kappa\rightarrow +0$.

 \noindent
 (ii) $R >  (N+1)(N+2)/2$\\
\noindent
 In this case, since the integrand in \eq{eq:gwithfresult} is oscillatory and has a modulus diverging in $r\rightarrow
 +\infty$,  the $\kappa\rightarrow +0$ limit has to be taken in a careful manner.
 For $\kappa\sim+0$, the integral is dominated by the large $r$ region.
 Therefore, the behavior of $g(N,R,\kappa)_{leading}$ in $\kappa\sim+0$ is given by
\[
g(N,R,\kappa)_{leading}&\sim
\kappa^{-\frac{\#P}{2}+\frac{NR}{6}}
\int^\infty dr\, r^{\gamma-1} e^{I \kappa^{1/3}r^2-\epsilon r^2}\CR
&\sim \kappa^{-\frac{\#P}{2}+\frac{NR}{6}}
\left(\epsilon-I \kappa^{\frac{1}{3}}\right)^{
-\frac{\gamma}{2}}.
\]
where $\gamma=NR-N(N+1)(N+2)/2$.
By taking the $\epsilon\rightarrow +0$ limit,
we find that 
\[
g(N,R,\kappa)_{leading}\sim\kappa^{0}.
\label{eq:caseii}
\]
Therefore $g(N,R,\kappa)$ converges to a finite value 
in the $\kappa\rightarrow +0$ limit.

\noindent
(iii) $R =  (N+1)(N+2)/2$\\
With slight modification of the discussions in the case (ii), we obtain
\[
g(N,R,\kappa)_{leading}\sim \log(\kappa).
\label{eq:caseiii}
\]
Therefore it diverges logarithmically.

Combining the three cases above, 
we see that the behavior of $g(N,R,\kappa)$ in $\kappa\rightarrow +0$ limit 
has a transition at $R=R_c:=(N+1)(N+2)/2$, 
where it is finite and diverging
at $R>R_c$ and $R\leq R_c$, respectively. 
However, this result may be changed by 
some corrections. 
As analyzed in Section~\ref{sub:AlphaLarg1}, 
there exists a transition region 
$R\sim N^{\alpha}\, (1\leq \alpha \leq 2)$,
and $R\sim R_c\sim N^2$ is 
at the edge of the transition 
region. 
Therefore,
$f_{N,R}(t)$
may 
have some important corrections at $R\sim R_c$
in addition to $f_{N,R}(t)_{leading}$. 
We also commented above that $f_{N,R}(t)$ may have some important corrections at 
$t\gtrsim N^2 R^2$.
In fact, the case (ii) has the main contributions 
from the large $t$ region if $\kappa$ is taken small.
Therefore, unless 
$\kappa$ is kept finite  
for $R\gg R_c$ in
the case (ii),
the results above must be taken with caution. 

Let us see the subtleties
more concretely by using our results, \eq{eq:allarger2} and \eq{eq:aleq2}, which are
from the analysis of the 
dominant graphs for $R\sim N^\alpha$ in large $N$
with $\alpha>2$ and $\alpha=2$, respectively.
By applying the relation \eq{eq:relofCandd} as before,
we obtain $f_{N,R}(t)_{leading}=\exp({\cal F}_{N,R}^{\rm dom}(\lambda,k))$
with the replacement $\lambda/k^3\rightarrow 8 t/N^3R^3$.  
In either of the cases \eq{eq:allarger2} and \eq{eq:aleq2},
$f_{N,R}(t)_{leading}$ at large $t$ has the divergence
coming from the second term,
which violates the monotonically decreasing property of $f_{N,R}(t)_{leading}$ discussed in Section~\ref{sec:probability}.
Therefore we encounter 
a maximum value of $t$, over which the expression of $f_{N,R}(t)_{leading}$ cannot be correct.
Then, the 
integration over $r$
cannot be done to the infinite,
and 
this is problematic in taking the $\kappa\rightarrow +0$ limit in the case (ii) and (iii),
since the main contribution
of the integration
comes from the large-$r$ region. 

From these discussions,
to obtain more concrete
statement in the $\kappa\rightarrow +0$ limit, we would need to 
check the sub-leading corrections to $f_{N,R}(t)$. 
On the other hand, we would be able to say that 
the present result is in favor of, or does not contradict,
the consistent
quantum probabilistic interpretation of the wave function of the tensor model
\cite{Obster:2017dhx}, because the tensor model requires $R=(N+2)(N+3)/2>R_c$,
and $g(N,R,\kappa)_{leading}$ 
is finite
in the $\kappa\rightarrow +0$ limit 
at least in the computation above.
To improve the statement,
in addition to computing the sub-leading corrections, one should
consider the absolute value
$|\psi(P)|^R$ as the integrand rather than the present $\psi(P)^R$ in \eq{eq:defofg}, since the former is 
semi-positive definite being a probability distribution, but the latter, the 
present one, is not.
Integrating over $P_{abc}$
in the former case 
leads to a more involved form
than \eq{eq:integral}, and analyzing it is 
left for future study.

\section{Summary and future prospects}

In this paper, we considered a random matrix model
with pairwise and non-pairwise contracted indices.
We analyzed the
model in
various regimes concerning the relative relation between $N$ and $R$, which are the  
dimensions of the pairwise and non-pairwise contracted indices, respectively.
We used Feynman diagrammatic expansions for the analysis, and have shown a transition of 
dominant graphs 
between tree-like 
ones at large $N$ and loop-like ones
at large $R$.
As a specific application, we applied our result to study the integrability 
of the wave function of the model
introduced in \cite{Obster:2017pdq} as a toy model 
for a tensor model
\cite{Ambjorn:1990ge,Sasakura:1990fs,Godfrey:1990dt} in the Hamilton formalism
\cite{Sasakura:2011sq,Sasakura:2012fb}.  
The result seems to be in favor of the consistency of the quantum probabilistic interpretation 
of the tensor model.

More precisely, in the regimes where $N$ is large but $R$ is finite, or where  $R\sim N^\alpha$ with $\alpha\le 1$, which includes the case of square matrices $\alpha=1$, we have shown the dominance of the tree-like graphs, which are the dominant graphs for the $\phi^6$ vector model. As explained in the Appendix~\ref{app:melo}, this tree-like family corresponds to the family of melonic graphs for the equivalent replicated vector model with random tensor couplings (or analogs with a time-dependence).  This shows the robustness of the dominance of the tree-like graphs for matrix models with non-pairwise contracted indices (such tree-like behaviors are also found for non-square one-matrix models) or of the dominance of melonic graphs for replicated random coupling vector models, when the number of replicas does not exceed the size of the system.

In the regime where $R\sim N^\alpha$ with $\alpha>1$, we have shown that the dominant graphs exhibit a very interesting behavior: tree-like graphs dominate for graphs with $\alpha/(\alpha - 1)$ or less interactions, while a family of very ordered star-like graphs dominate for graphs with more interactions. In a sense, the small dominant graphs exhibit ``more disorder'' than the larger ones.  The contribution of the tree-like graphs is a truncation of the usual vector-model free-energy, and the contributions of the star-like graphs is roughly a logarithm.   The value of $\alpha$ thus provides a parameter to tune the value at which the vector-model free-energy is truncated, and the remainder replaced by a logarithm expansion.  While such an interesting behavior is not known to the authors to exist in other vector, matrix, or tensor models, it is not possible to rescale the coupling constants to obtain a finite limit for the free-energy that involves both families, precisely because of the different behaviors in $N$ and $R$ of the tree-like and star-like graphs. It would be interesting to see if a similar situation occurs for dominant graphs for other models with non-pairwise contracted indices. Our first guess is that we could have such a competition between necklace-like graphs and tree-like graphs whenever an odd number of indices are contracted together, while trees could dominate for any $\alpha$ for an even number of contracted indices, however the precise situation when $\alpha$ takes values in $\bR^+$ should be investigated. 

It should be stressed that we have not concluded in the present paper that the  wave function of the model introduced in \cite{Obster:2017pdq} is integrable in the region $R\sim N^2$, which is the region of interest for the the tensor model of \cite{Sasakura:2011sq,Sasakura:2012fb}. 
This is due to some limitations of our approximation at the regime $R\sim N^2$, which 
comes from the condition  $R=(N+2)(N+3)/2$ required for the consistency \cite{Sasakura:2013wza} of the tensor model of \cite{Sasakura:2011sq,Sasakura:2012fb}.
Therefore an obviously important question about our results is how they would change  by improving the approximations.
This could be done by including higher order Feynman graphs along the present line, or employing 
the various techniques which have been developed for the analysis of the spin glass  models, because of the similarities between our model and the $p$-spin spherical model \cite{pspin,pedestrians}.
In particular, it is important to obtain
more correctly the behavior of  the moment-generating function $f_{N,R}(t)$, especially in the $t\rightarrow \infty$ limit,
because it essentially determines  whether the wave function of the toy model \cite{Obster:2017pdq}  related  to the tensor model is integrable or not. It is also important to deal with the real wave function of the tensor model, which has the same interaction term, but for which the Gaussian part is replaced by a product of Airy functions \cite{Narain:2014cya,Obster:2017dhx}.

An interesting future question is to find a way to take a large $N$ limit which would keep 
the non-trivial characteristics of this model. 
We have observed that there is a crossover between the tree-like and the star-like graphs under 
varying the parameter $\alpha$ of $R\sim N^\alpha$. 
However, as pointed out in the text, it is not possible at the level of the Feynman graph series expansion to take a large $N$ limit which keeps both the tree and star graphs in an interesting manner.
On the other hand, a close look at \eq{eq:freeenergylargeR} suggests an interesting scenario.
To see this, let us consider 
${\cal F}_{N,R}(\lambda,k)/RN$,
which is the free energy per degrees of freedom, and take a large $N$ 
limit with 
$\lambda/k^3 \sim N^{-2}$ and $R\sim N^2$.
The former scaling of the coupling constants ensures the tree-like graphs to give 
a non-trivial contribution to  ${\cal F}_{N,R}/RN$ at large-$N$.
Assuming that the contribution to ${\cal F}_{N,R}/RN$
for the star graphs actually 
vanishes in this limit\footnote{As mentioned in the text,  we cannot rely on the logarithm expression for the first term in \eq{eq:freeenergylargeR} to draw conclusions in this regime on the relative contributions of the star-like graphs with respect to tree-like graphs or other necklace graphs, as the series for star-like graphs is highly divergent with these choices for the coupling constants, while the series for trees and necklaces are convergent. },
the second term, which contains the other {\rm necklace} graphs,
remains finite and non-trivial. 
Note that this scenario does not contradict our analysis in the text,
since order by order analysis of the dominant graphs 
is not necessarily related to the dynamics of the model in general.
At the present stage, this scenario is not more than a 
speculation, since we presently do not understand well the dynamics at $R\sim N^2$.

 \vspace{0.8cm}
\section*{Acknowledgements}
The work of N.S. is supported in part by JSPS KAKENHI Grant No.15K05050. 
L.L. is a JSPS international research fellow.
N.S. would like to thank S.~Nishigaki for brief communication,
and G.~Narain for some inspiring comments.
L.L. would like to thank V.~Bonzom for useful discussions, as well as L.~Cugliandolo for useful references. 

\appendix
\section*{Appendices}

\section{Tree dominance at large $N$ and Schwinger-Dyson equation}
\label{app:proof}
In this appendix, we first prove the dominance of tree-like graphs in the large $N$ limit, and then
write down the self-consistency equation (Schwinger-Dyson equation) for the two-point function. 
The explanation of these matters was kept brief in the text, because they are assumed to be well-known standard knowledge.
This appendix could be useful for readers who are not familiar with these results or the way we have  formulated them.

\begin{figure}
    \centering
    \includegraphics[width=2.8cm]{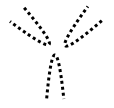}
    \hfil
    \includegraphics[width=2.25cm]{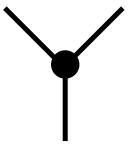}
    \caption{Left: An interaction represented only with the dotted edges. Right: The simplified representation
    of the interaction.}
    \label{fig:dotted}
\end{figure}

\begin{figure}
 \begin{center}
 \includegraphics[width=5cm]{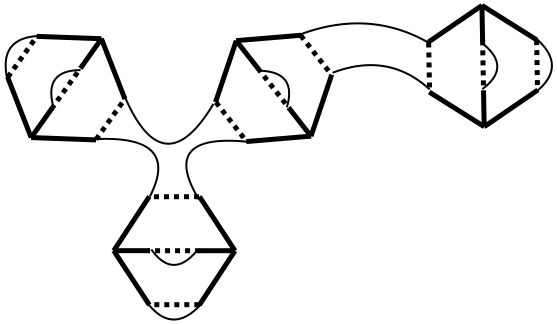}
\hfil
 \includegraphics[width=4.5cm]{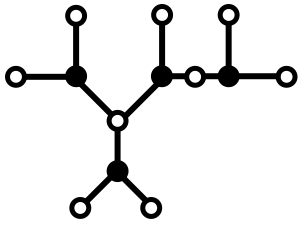}
 \caption{Left: An example of a tree-like graph. 
 Right: The same graph represented using the simplified
 vertices. Each white blob corresponds to a dotted loop,
 showing that this graph has weight $N^9$.}
 \label{fig:largeNexsimple}
 \end{center}
 \end{figure}

In the case where $N$ is large but $R$ is finite, what matters is the weight in $N$, namely, the number of dotted 
loops (faces) in a graph,
which correspond to free sums over the lower indices of $\phi$.
One can thus ignore the solid lines, which correspond to the upper indices.
This means that the interaction in Figure~\ref{fig:intvertex} can be simplified to 
an interaction with only dotted edges as on the left of Figure~\ref{fig:dotted} (this representation is common for vector models) or even simply by a black blob with three legs attached, as on the right of this figure, as will be explained in detail below.
The graph on the right of Figure~\ref{fig:largeNexsimple} gives an example of a graph represented by such simplified vertices. This example corresponds to the graph on the left of Figure~\ref{fig:largeNexsimple},
which uses the usual representation of the interaction (Figure~\ref{fig:intvertex}).
In the simplified graphs,  the connecting points among the legs of the simplified vertices are represented by white blobs.
Each white blob corresponds to a dotted loop, and therefore
the weight in $N$ of a graph is the number of white
blobs in it. 
While the vertices corresponding to the interactions are trivalent, the number of legs attached to each white blob is not restricted.
 
 Let us consider a connected graph  constructed in the above way with  $n$ simplified vertices.
Its first Betti number $B_1$  is given by
 \[
B_1=3n-\left(n+\sum_{i\ge 1} b_i\right)+1,
\label{eq:largeNeqs}
\]
where $b_i$ denotes the number of white blobs which have $i$ legs attached.
Here $3n$ is the total number of edges and the quantity in the parenthesis is the total number of vertices, namely, white and black blobs. 
Since the weight in $N$ is $\Fd=\sum_{i\ge1} b_i$,
we obtain from \eq{eq:largeNeqs}
\be
\Fd= 2n+1-B_1.
\ee
Since $B_1\in \mathbb{N}$, $F_d$ takes the maximal value $2n+1$ for $B_1=0$, namely, 
if and only if the graph is a tree. 

For a tree, it is easy to prove that the 
weight in $R$ is $R$.
Therefore, a tree graph has the weight $N^{2n+1}R$ 
in $N$ and $R$.
 
From the above proof, in the leading order of $N$, the two-point correlation function $\langle \phi_a^i \phi_b^j \rangle$ can be computed
by summing over all the graphs
which are obtained by opening any of the Wick contraction edges (the thin edges) in the tree 
vacuum graphs. 
Then, it is easy to see that the two-point
correlation function necessarily has the form,
\[
\langle \phi_a^i \phi_b^j \rangle_{tree}=G \delta_{ab}\delta^{ij},
\label{eq:defofG}
\] 
where $G$ depends on $\lambda,k$ and $N$.

\begin{figure}
  \begin{center}
 \includegraphics[width=10cm]{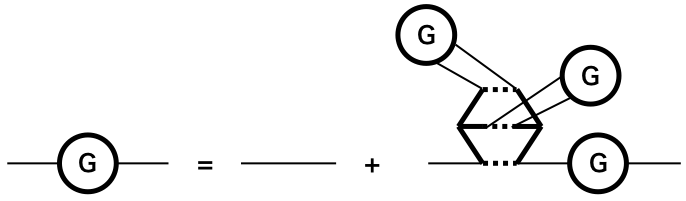}
 \caption{The Schwinger-Dyson equation for the two-point
 correlation function.}
 \label{fig:SD}
 \end{center}
 \end{figure}
From a consistency equation for the tree graphs,  
one can obtain the Schwinger-Dyson equation shown in 
Figure~\ref{fig:SD}.
By putting the form \eq{eq:defofG}, this leads to 
\be
G=\frac{1}{2k}-\frac{6\lambda N^2}{2k} G^3.
\label{eq:SD}
\ee
After a rescaling, one obtains \eq{eq:Ternary-Tree}.

\section{Feynman graphs of the replicated vector model with random tensor couplings and melonic graphs}
\label{app:melo}
\begin{figure}[h!]
\center
\includegraphics[scale=0.8]{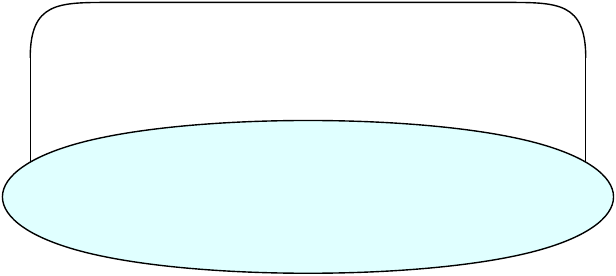}
\hspace{1cm}\raisebox{1cm}{$\rightarrow$}\hspace{1cm}
\includegraphics[scale=0.8]{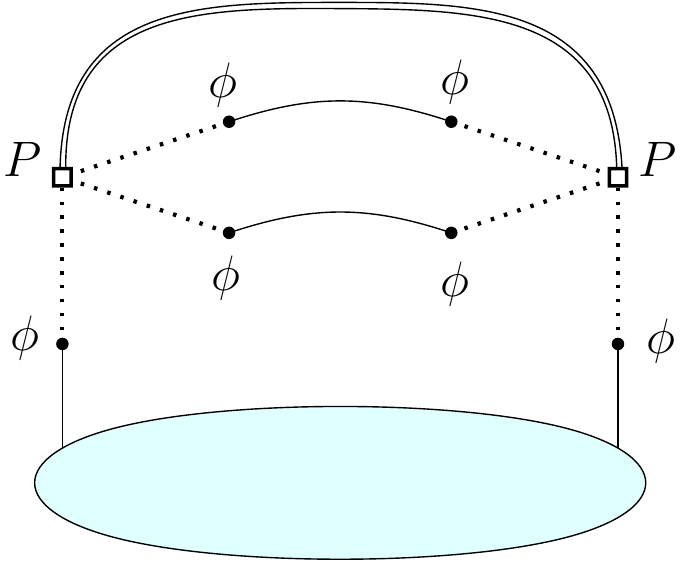}
\caption{Recursive contruction of a melonic graph for $R=1$. The double lines represent the Wick contraction of the two tensors $P$ in the pair.}
\label{fig:MeloR1}
\end{figure}

We consider the following expression, 
\[
Z_{N,R}(\lambda, k) = \int dP e^{-\frac 1 2 \sum_{abc=1}^N P_{abc}^2}\Bigl(\int d\phi e^{ - k \sum_{a=1}^N \phi_a^2  - I\sqrt{2\lambda } \sum_{a,b,c=1}^N P_{abc} \phi_a \phi_b \phi_c }\Bigr)^R.
\] 
Labeling the $R$ copies of $\phi$ from 1 to $R$, we obtain 
\[
\label{eqref:random-coup}
Z_{N,R}(\lambda, k) = \int dP \int d\phi\ e^{-\frac 1 2 \sum_{abc=1}^N P_{abc}^2}e^{ - k \sum_{a=1}^N \sum_{i=1}^R\phi_a^i  - I\sqrt{2\lambda } \sum_{a,b,c=1}^N P_{abc} (\sum_{i=1}^R\phi^i_a \phi^i_b \phi^i_c) }.
\] 
The initial model \eqref{eq:integral} is recovered after integrating over $P$. Let us take a closer look at the Feynman graphs of \eqref{eqref:random-coup}. In the case where $R=1$, we represent each tensor $P$ by a vertex with three dotted edges attached, one for each one of the vectors. At the end of the dotted edge is a vertex representing the vector $\phi$. The integration over $\phi$ generates trivalent graphs\footnote{More precisely, in the convention we use, it generates Wick pairings between the $\phi$, represented as thin edges between the corresponding vertices, so that two trivalent vertices corresponding to tensors $P$ are linked by chains formed by a dotted edge, a thin edge, and another dotted edge.} whose vertices carry tensors $P$, and to which are associated polynomials in $P$ obtained by contracting the indices according to the edges in the graphs. For a given graph, the integration over $P$ is expressed as a sum over Wick pairings of the tensors, along with a power of $N$ corresponding to the number of faces thus created. 
It is well known that at large $N$, the  graphs that dominate are the so-called {\it melonic graphs } obtained by recursively adding pairs of vertices as in Fig.~\ref{fig:MeloR1}, \cite{melons1, melons2, melons3}. The dominance of these graphs is one of the reasons of the success of the SYK model \cite{SYKSY,SYKK} and related models: while these graphs have a simple recursive structure so that the theory is exactly solvable in the IR, they are not trivial and contain very interesting physics. 

\medskip

\begin{figure}[h!]
\center
\includegraphics[scale=0.8]{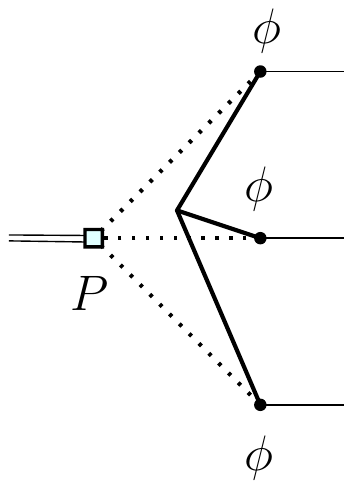}
\hspace{1cm}\raisebox{2cm}{;}\hspace{1cm}
\raisebox{0.5cm}{\includegraphics[scale=0.6]{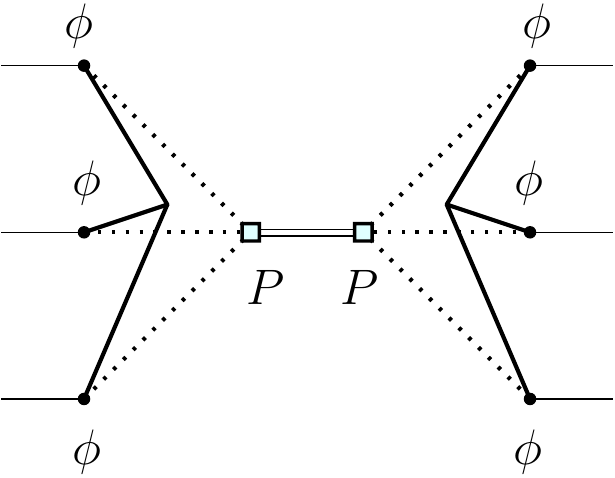}}
\hspace{0.5cm}\raisebox{2cm}{$\rightarrow$}\hspace{0.5cm}
\raisebox{0.5cm}{\includegraphics[scale=0.6]{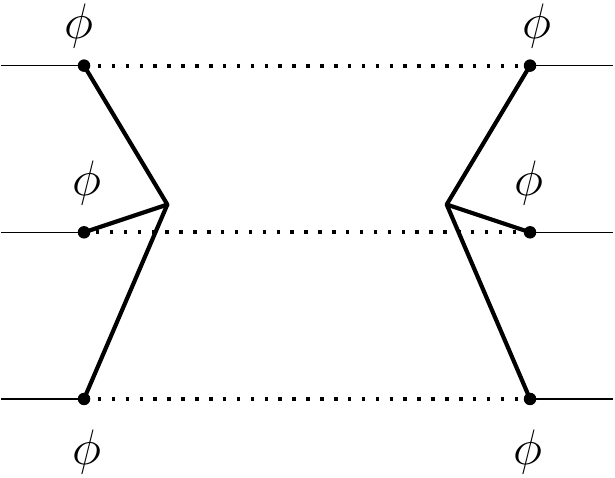}}
\caption{On the left is the interaction for the replicated model with random couplings. The way the interaction of the matrix model \eqref{eq:integral} is recovered by integrating the $P$  is shown on the right.}
\label{fig:Int-rand-coup}
\end{figure}

We now return to the case where $R>1$. The vertices corresponding to the $\phi$ are then linked by new solid edges that meet at trivalent nodes (Fig.~\ref{fig:Int-rand-coup}). The graphs representing the interactions of our model (Fig.~\ref{fig:intvertex}) are obtained by deleting the edges corresponding to the Wick pairings between the $P$, as well as the vertices corresponding to the $P$, and identifying the dotted half-edges, as illustrated on the right of  Fig.~\ref{fig:Int-rand-coup}.

\medskip

\begin{figure}[h!]
\center
\includegraphics[scale=0.8]{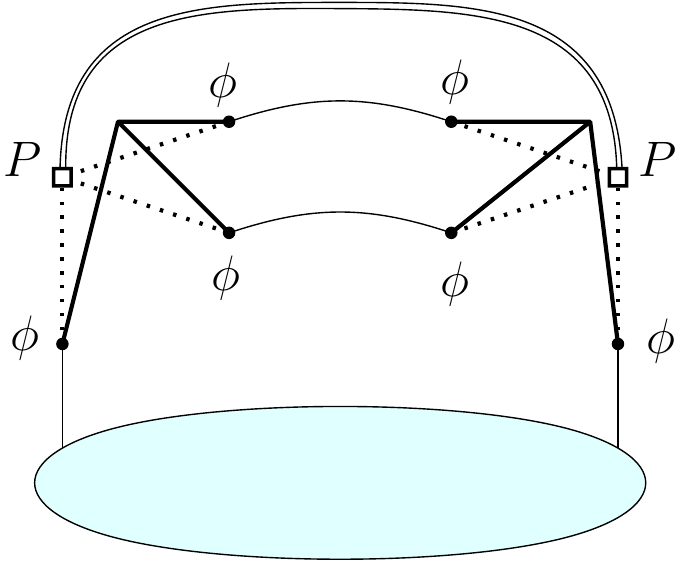}
\hspace{1cm}\raisebox{1cm}{$\Leftrightarrow$}\hspace{1cm}
\includegraphics[scale=0.8]{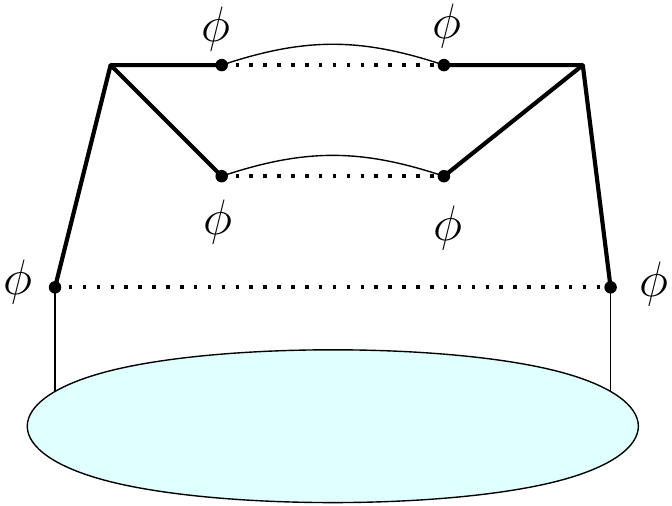}
\caption{The melonic graphs in the replicated model with random couplings correspond to the tree-like graphs for the model \eqref{eq:integral}.}
\label{fig:Melo-eq-tree}
\end{figure}

As shown in Fig.~\ref{fig:Melo-eq-tree}, the tree-like graphs for the model studied in the present paper precisely correspond to the melonic graphs for the replicated model with random tensor couplings. Therefore, as explained in Section \ref{sub:LargeNfiniteR}, the dominance of the melonic graphs still holds at large $N$ for the model with real replicas, i.e.~when $R$ is finite \cite{MezPar, pedestrians}.  This dominance is quite robust, as we showed that it actually still holds when the number $R$ of replicas is large, but smaller or equal to the size $N$ of the system ($R\sim N^\alpha$ with $0<\alpha\le 1$), although we do not know of any statistical physics interpretation for such a regime. This should still be true when the fields $\phi$ have a time dependence.

\medskip

\section{Explicit expression for the radial integration
}
\label{app:explicit}
 In this appendix, we will show 
\[
\int_0^\infty dr\, r^{NR-1}   e^{-\lambda r^6-k r^2}&=
\frac{\lambda^{-\frac{NR}{6}}\,\Gamma\left(\frac{NR}{6}\right)}{6\,\Gamma(1)}
{}_1F_{2}\left(\frac{NR}{6};\frac{1}{3},\frac{2}{3};-\frac{k^3}{27\lambda}\right)\CR
&-
\frac{\lambda^{-\frac{NR}{6}-\frac{1}{3}}k\,\Gamma\left(\frac{NR}{6}+\frac{1}{3}\right)}{6\,\Gamma(2)}
{}_1F_{2}\left(\frac{NR}{6}+\frac{1}{3};\frac{2}{3},\frac{4}{3};-\frac{k^3}{27\lambda}\right)\CR
&+
\frac{\lambda^{-\frac{NR}{6}-\frac{2}{3}}k^2\,\Gamma\left(\frac{NR}{6}+\frac{2}{3}\right)}{6\,\Gamma(3)}
{}_1F_{2}\left(\frac{NR}{6}+\frac{2}{3};\frac{4}{3},\frac{5}{3};-\frac{k^3}{27\lambda}\right),
\label{eq:explicitofintegral}
\]
 where the generalized hypergeometric function ${}_1F_2(a;b_1,b_2;z)\ (-b_i \notin \mathbb{N})$ is defined by
 \[
 {}_1F_2(a;b_1,b_2;z)=\sum_{n=0}^\infty \frac{(a)_n}{(b_1)_n (b_2)_n}
 \frac{z^n}{n!}
 \label{eq:1f2}
 \]  
 with the Pochhammer symbol,
 \[
 (x)_n=
 \left\{
 \begin{array}{cl}
1 & n=0 ,\\
x(x+1)\cdots(x+n-1) & n>0.
 \end{array}
 \right.
 \]
Since the series \eq{eq:1f2} converges everywhere in $z$, ${}_1F_2(a;b_1,b_2;z)$ is an entire function of $z$.
It has an essential singularity at $z=\infty$ for $-a\notin \mathbb{N}$, since
the expansion \eq{eq:1f2} of this entire function is an infinite series.
Therefore \eq{eq:explicitofintegral} has an essential singularity at $\lambda=0$.
 
 To show \eq{eq:explicitofintegral}, it is enough to
 employ the following convergent series representing
 an integral:
 \[
\int_0^\infty dr\, r^{N-1}  e^{-r^6-h r^2}&=\sum_{n=0}^\infty \frac{(-h)^n}{n!}
\int_0^\infty dr\, r^{N-1+2n} e^{-r^6}\CR
&=\frac{1}{6}\sum_{n=0}^\infty \frac{(-h)^n}{n!}\Gamma\left(\frac{N}{6}+\frac{n}{3}\right).
\]
By dividing the $n$ summation into the three cases, $n=3m,3m+1,3m+2\ (m=0,1,\ldots)$
and changing the variables appropriately,  we obtain \eq{eq:explicitofintegral}.


\begin{thebibliography}{10}

\bibitem{Wigner} E.~Wigner, ``Characteristic vectors of bordered matrices with infinite dimensions'', Annals of Mathematics {\bf 62} (3): 548-564.

  \bibitem{thooft-planar}
  G.~'t Hooft, ``A planar diagram theory for strong interactions,'' Nucl.\ Phys.\ B {\bf 72}, 461 (1974).
  
\bibitem{Brezin:1990rb} 
  E.~Brezin and V.~A.~Kazakov,
  ``Exactly Solvable Field Theories of Closed Strings,''
  Phys.\ Lett.\ B {\bf 236}, 144 (1990).
  doi:10.1016/0370-2693(90)90818-Q
  
\bibitem{Douglas:1989ve} 
  M.~R.~Douglas and S.~H.~Shenker,
  ``Strings in Less Than One-Dimension,''
  Nucl.\ Phys.\ B {\bf 335}, 635 (1990).
  doi:10.1016/0550-3213(90)90522-F
  
\bibitem{Gross:1989vs} 
  D.~J.~Gross and A.~A.~Migdal,
  ``Nonperturbative Two-Dimensional Quantum Gravity,''
  Phys.\ Rev.\ Lett.\  {\bf 64}, 127 (1990).
  doi:10.1103/PhysRevLett.64.127
  
  \bibitem{2DgravityReview}
  P.~Di Francesco, P.~H.~Ginsparg and J.~Zinn-Justin, ``2-D Gravity and random matrices,''
Phys.\ Rept.\ {\bf 254}, 1 (1995)

  \bibitem{oxford-rand-mat}
  G.~Akemann, J.~Baik and P.~Di Francesco, ``The Oxford Handbook of Random Matrix
Theory'', Oxford University Press.

  
  \bibitem{book-rand-mat-2}
  G.~W.~Anderson, A.~Guionnet and O.~Zeitouni, ``An Introduction to Random Matrices'', Studies in Advanced Mathematics, v.\ 118, Cambridge University Press (2009).
  
\bibitem{Anderson:1990nw} 
  A.~Anderson, R.~C.~Myers and V.~Periwal,
  ``Complex random surfaces,''
  Phys.\ Lett.\ B {\bf 254}, 89 (1991).
  

  
\bibitem{Anderson:1991ku} 
  A.~Anderson, R.~C.~Myers and V.~Periwal,
  ``Branched polymers from a double scaling limit of matrix models,''
  Nucl.\ Phys.\ B {\bf 360}, 463 (1991).
  
\bibitem{Myers:1992dq} 
  R.~C.~Myers and V.~Periwal,
  ``From polymers to quantum gravity: Triple scaling in rectangular random matrix models,''
  Nucl.\ Phys.\ B {\bf 390}, 716 (1993)
  [hep-th/9112037].
  
  \bibitem{Francesco-rect}
    P.~Di Francesco,
    ``Rectangular Matrix Models and Combinatorics of Colored Graphs",
  Nucl.\ Phys.\ B {\bf 648} (2003) 461-496.
  
  
  %

  
\bibitem{Nishigaki:1990sk} 
  S.~Nishigaki and T.~Yoneya,
  ``A nonperturbative theory of randomly branching chains,''
  Nucl.\ Phys.\ B {\bf 348}, 787 (1991).
  doi:10.1016/0550-3213(91)90215-J
  
\bibitem{DiVecchia:1991vu} 
  P.~Di Vecchia, M.~Kato and N.~Ohta,
  ``Double scaling limit in O(N) vector models in D-dimensions,''
  Int.\ J.\ Mod.\ Phys.\ A {\bf 7}, 1391 (1992).
  doi:10.1142/S0217751X92000612
  
\bibitem{pspin}
A.~Crisanti and H.-J.~Sommers,
``The spherical p-spin interaction spin glass model: the statics,''
Z.\ Phys.\ B {\bf 87}, 341 (1992).
  
\bibitem{pedestrians}
T.~Castellani and A.~Cavagna, ``Spin-glass theory for pedestrians'', 
J.~Stat.~Mech.: Theo.~Exp. {\bf 2005}, 05012
[arXiv: cond-mat/0505032].

\bibitem{MezPar}
M.~M\'ezard, G.~Parisi,
``A first-principle computation of the thermodynamics of glasses",
J.\ Chem.\ Phys.\ {\bf 111}, 1076 (1999).
        
\bibitem{Narain:2014cya} 
  G.~Narain, N.~Sasakura and Y.~Sato,
  ``Physical states in the canonical tensor model from the perspective of random tensor networks,''
  JHEP {\bf 1501}, 010 (2015)
  doi:10.1007/JHEP01(2015)010
  [arXiv:1410.2683 [hep-th]].
          
        
\bibitem{Obster:2017dhx} 
  D.~Obster and N.~Sasakura,
  ``Emergent symmetries in the canonical tensor model,''
  PTEP {\bf 2018}, no. 4, 043A01 (2018)
  doi:10.1093/ptep/pty038
  [arXiv:1710.07449 [hep-th]].
  
\bibitem{Ambjorn:1990ge}
J.~Ambjorn, B.~Durhuus, and T.~Jonsson, ``{Three-dimensional simplicial quantum
  gravity and generalized matrix models},''
\href{http://dx.doi.org/10.1142/S0217732391001184}{{\em Mod. Phys. Lett.}
  {\bfseries A06} (1991) 1133--1146}.

\bibitem{Sasakura:1990fs}
N.~Sasakura, ``{Tensor model for gravity and orientability of manifold},''
\href{http://dx.doi.org/10.1142/S0217732391003055}{{\em Mod. Phys. Lett.}
  {\bfseries A06} (1991) 2613--2624}.

\bibitem{Godfrey:1990dt}
N.~Godfrey and M.~Gross, ``{Simplicial quantum gravity in more than
  two-dimensions},''
\href{http://dx.doi.org/10.1103/PhysRevD.43.1749}{{\em Phys. Rev.} {\bfseries
  D43} (1991) R1749--1753}.

  
  
  
\bibitem{Sasakura:2011sq} 
  N.~Sasakura,
  ``Canonical tensor models with local time,''
  Int.\ J.\ Mod.\ Phys.\ A {\bf 27}, 1250020 (2012)
  doi:10.1142/S0217751X12500200
  [arXiv:1111.2790 [hep-th]].
  
\bibitem{Sasakura:2012fb} 
  N.~Sasakura,
  ``Uniqueness of canonical tensor model with local time,''
  Int.\ J.\ Mod.\ Phys.\ A {\bf 27}, 1250096 (2012)
  doi:10.1142/S0217751X12500960
  [arXiv:1203.0421 [hep-th]].
  
 
\bibitem{Obster:2017pdq} 
  D.~Obster and N.~Sasakura,
  ``Symmetric configurations highlighted by collective quantum coherence,''
  Eur.\ Phys.\ J.\ C {\bf 77}, no. 11, 783 (2017)
  doi:10.1140/epjc/s10052-017-5355-y
  [arXiv:1704.02113 [hep-th]].
  
  
\bibitem{Sasakura:2014zwa} 
  N.~Sasakura and Y.~Sato,
  ``Ising model on random networks and the canonical tensor model,''
  PTEP {\bf 2014}, no. 5, 053B03 (2014)
  doi:10.1093/ptep/ptu049
  [arXiv:1401.7806 [hep-th]].
  
\bibitem{Sasakura:2014yoa} 
  N.~Sasakura and Y.~Sato,
  ``Exact Free Energies of Statistical Systems on Random Networks,''
  SIGMA {\bf 10}, 087 (2014)
  doi:10.3842/SIGMA.2014.087
  [arXiv:1402.0740 [hep-th]].
  
 \bibitem{SYKSY} S.~Sachdev and J.~Ye, ``Gapless spin fluid ground state in a random, quantum Heisenberg
magnet," Phys. Rev. Lett. {\bf 70} ,3339 (1993)  [arXiv:cond-mat/9212030].

\bibitem{SYKK} A.~Kitaev, ``A simple model of quantum holography", 
Talks at KITP, April 7 \url{http://online.kitp.ucsb.edu/online/entangled15/kitaev/} and May 27, 2015 \url{http://online.kitp.ucsb.edu/online/entangled15/kitaev2/}.

\bibitem{melons1}
V.~Bonzom, R.~Gurau, A.~Riello, and V.~Rivasseau, ``Critical behavior of colored tensor models in the large $N$ limit'', Nuclear Physics B {\bf 853} pp.174-195, [arXiv:1105.3122].

\bibitem{melons2} V.~Bonzom, R.~Gurau, and V.~Rivasseau, ``Random tensor models in the large $N$ limit: Uncoloring the colored tensor models,'' Phys.\ Rev.\ D {\bf 85}, 084037.

\bibitem{melons3} V.~Bonzom, V.~Nador and A.~Tanasa, ``Diagrammatic proof of the large N melonic dominance in the SYK model,'' arXiv:1808.10314 [math-ph].

\bibitem{SAPM:SAPM192761164}
F.~L. Hitchcock, ``The expression of a tensor or a polyadic as a sum of
  products,'' \href{http://dx.doi.org/10.1002/sapm192761164}{{\em Journal of
  Mathematics and Physics} {\bfseries 6} no.~1-4, (1927) 164--189}.
  \url{http://dx.doi.org/10.1002/sapm192761164}.

\bibitem{comon:hal-00923279}
P.~Comon, ``{Tensors: a Brief Introduction},''
  \href{http://dx.doi.org/10.1109/MSP.2014.2298533}{{\em {IEEE Signal
  Processing Magazine}} {\bfseries 31} no.~3, (May, 2014) 44--53}.
  \url{https://hal.archives-ouvertes.fr/hal-00923279}.

\bibitem{Landsberg2012}
{J.~M.~Landsberg}, {\em {Tensors: Geometry and Applications}},
\newblock {American Mathematical Society, Providence}, {2012}.

\bibitem{Sasakura:2013wza} 
  N.~Sasakura,
  ``Quantum canonical tensor model and an exact wave function,''
  Int.\ J.\ Mod.\ Phys.\ A {\bf 28}, 1350111 (2013)
  doi:10.1142/S0217751X1350111X
  [arXiv:1305.6389 [hep-th]].

\end{thebibliography}
 
\providecommand{\href}[2]{#2}\begingroup\raggedright\endgroup

\end{document}